\documentclass[aps,prl,twocolumn,showpacs,groupedaddress]{revtex4}  
\usepackage{graphicx}  
\usepackage{dcolumn}   
\usepackage{bm}        
\usepackage{amssymb}   

\usepackage{subfigure}

\newcommand{\ppb}{\mbox{\ensuremath{p\bar p}}}

\newcommand{\ttb}{\mbox{\ensuremath{t\bar t}}}
\newcommand{\bbb}{\mbox{\ensuremath{b\bar b}}}
\newcommand{\nunub}{\mbox{\ensuremath{\nu\bar\nu}}}

\newcommand{\invfb}{fb$^{-1}$}
\newcommand{\Gev}{GeV}

\newcommand{\met}{\mbox{\ensuremath{\not\!\!E_T}}}
\newcommand{\mht}{\mbox{\ensuremath{\not\!\!H_T}}}
\newcommand{\mpt}{\mbox{\ensuremath{\slash\kern-.5emp_{T}}}}
\newcommand{\et}{\mbox{\ensuremath{E_{T}}}}

\newcommand{\hht}{\mbox{\ensuremath{H_{T}}}}
\newcommand{\pt}{\mbox{\ensuremath{p_{T}}}}

\newcommand{\jj}{\mbox{\ensuremath{(\mathrm{jet}_1,\mathrm{jet}_2)}}}

\begin{document}



\hspace{5.2in} \mbox{FERMILAB-PUB-09-649-E-PPD}

\title{Search for the Standard Model Higgs Boson in the 
{\boldmath $ZH\to\nunub\bbb$} Channel \\ 
in 5.2~fb{\boldmath $^{-1}$} of {\boldmath $p\bar{p}$} Collisions at 
{\boldmath $\sqrt{s}=1.96$}~TeV 
}
%
\author{V.M.~Abazov$^{37}$}
\author{B.~Abbott$^{75}$}
\author{M.~Abolins$^{64}$}
\author{B.S.~Acharya$^{30}$}
\author{M.~Adams$^{50}$}
\author{T.~Adams$^{48}$}
\author{E.~Aguilo$^{6}$}
\author{G.D.~Alexeev$^{37}$}
\author{G.~Alkhazov$^{41}$}
\author{A.~Alton$^{64,a}$}
% ...

\affiliation{\vspace{0.1 in}{\rm (D0 Collaboration)}\vspace{0.1 in}}
% ...
\affiliation{$^{6}$Simon Fraser University, Burnaby, British Columbia, Canada;
                and York University, Toronto, Ontario, Canada}
% ...
\affiliation{$^{30}$Tata Institute of Fundamental Research, Mumbai, India}
% ...
\affiliation{$^{37}$Joint Institute for Nuclear Research, Dubna, Russia}
% ...
\affiliation{$^{41}$Petersburg Nuclear Physics Institute,
                St. Petersburg, Russia}
% ...
\affiliation{$^{48}$Florida State University, Tallahassee, Florida 32306, USA}
% ...
\affiliation{$^{50}$University of Illinois at Chicago,
                Chicago, Illinois 60607, USA}
% ...
\affiliation{$^{64}$Michigan State University,
                East Lansing, Michigan 48824, USA}
% ...
\affiliation{$^{75}$University of Oklahoma, Norman, Oklahoma 73019, USA}
% ...

           
\begin{abstract}
A search is performed for the standard model Higgs boson in 5.2~\invfb\ 
of \ppb\ collisions at $\sqrt{s}=1.96$~TeV, 
collected with the D0 detector at the Fermilab Tevatron Collider. 
The final state considered is a pair of $b$ jets and large
missing transverse energy, 
as expected from $\ppb\to ZH\to\nunub\bbb$ production. The search is also 
sensitive to the $WH\to\ell\nu\bbb$ channel when the charged lepton is not 
identified. For a Higgs boson mass of 115~GeV, a limit is set at the 
95\% C.L. on 
the cross section multiplied by branching fraction for 
$[\ppb\to (Z/W)H](H\to\bbb)$  
that is a factor of 3.7 larger than the standard model value, 
consistent with the factor of 4.6 expected. 
\end{abstract}

\maketitle


The existence of the Higgs boson is the only fundamental 
element of the standard model 
(SM) that has yet to be confirmed. Its observation would be a key step in 
establishing the mechanism
of electroweak symmetry breaking and mass generation.
Associated $ZH$ production in \ppb\ collisions, 
with $Z\to\nunub$ and $H\to\bbb$, is among the most
sensitive processes for seeking  
a Higgs boson with a mass $m_H \lesssim 135$~GeV at the Fermilab 
Tevatron Collider~\cite{prospects}.
The D0 Collaboration published a search for this process based on 
0.9~\invfb\ of integrated luminosity~\cite{theo}. 
The CDF Collaboration recently released the results of a search
using 2.1~\invfb~\cite{cdfbbmet}. 
A lower limit of 114.4~GeV was set
by the LEP experiments on the mass of the Higgs boson 
from searches for the reaction
$e^+e^-\to ZH$~\cite{LEPH}, while an indirect
upper limit of 157~GeV can be inferred 
from precision electroweak data~\cite{EW}. These limits and those given below
are all defined at the 95\% C.L. 

This Letter presents a new search using an integrated luminosity more than
5 times larger than in~\cite{theo}. 
The final-state topology considered consists of a pair of 
$b$ jets from $H\to\bbb$ and missing transverse energy (\met) 
from $Z\to\nunub$.
The search is therefore also 
sensitive to the $WH$ process when the charged lepton 
from $W\to\ell\nu$ decay is not identified. The main backgrounds arise from 
$(W/Z)$+heavy flavor jets (jets initiated by $b$ and $c$ quarks), 
top quark production, and multijet (MJ) events 
with \met\ arising from mismeasurement of jet energies.


The D0 detector is described in~\cite{Dzero}.
The data used in this analysis were recorded using 
triggers designed to select events with jets and \met~\cite{theo,ochando}.
After imposing data quality requirements, 
the total integrated luminosity~\cite{lumi} 
is 5.2~\invfb. 
The analysis relies on  
(i) charged particle tracks, (ii) calorimeter jets 
reconstructed in a cone of radius 0.5, using the iterative midpoint cone 
algorithm~\cite{jetalgo}, and (iii)
electrons or muons identified through the association
of tracks 
with electromagnetic calorimeter clusters or with hits in 
the muon detector, respectively.
The \met\ is reconstructed as the opposite of the vectorial sum of
transverse components of energy deposits in the calorimeter
and is corrected for identified muons. 
Jet energies are calibrated using transverse energy balance
in photon+jet events~\cite{mikko}, 
and these corrections are propagated to the \met.

Backgrounds from SM processes are determined through
Monte Carlo simulation, while instrumental MJ background 
is estimated from data. 
Events from $(W/Z)$+jets processes are generated with 
{\sc alpgen}~\cite{alpgen}, interfaced with {\sc pythia}~\cite{pythia} for 
initial and final-state radiation and for hadronization. 
The \pt\ spectrum of the $Z$ is reweighted to match the D0 
measurement~\cite{Zpt}. The \pt\ spectrum of the $W$ is reweighted 
using the same experimental input, corrected for the differences between 
the $Z$ and $W$ \pt\ spectra predicted in next-to-next-to-leading order 
(NNLO) QCD~\cite{MelPet}.
For \ttb\ and electroweak single top quark production, the {\sc alpgen} and 
{\sc comphep}~\cite{comphep} generators, respectively, 
are interfaced with {\sc pythia}, while 
vector boson pair production is generated with {\sc pythia}. The 
$ZH$ and $WH$ signal processes are generated with {\sc pythia} 
for Higgs boson masses ($m_H$) from 100 to 150 GeV, in 5 GeV steps. 
All these simulations use CTEQ6L1 parton distribution functions 
(PDFs)~\cite{cteq}.

The absolute normalizations for $(W/Z)$+jets production are obtained from
NNLO calculations of
total cross sections based on~\cite{WZcross}, 
using the MRST2004 NNLO PDFs~\cite{MRST}.
The heavy-flavor fractions are obtained using {\sc mcfm}~\cite{mcfm}. 
Cross sections for
other SM backgrounds are taken from~\cite{xsections}, or
calculated with {\sc mcfm}, and  
the cross sections for signal are taken from~\cite{signal}.

Signal and background samples are passed through a full 
{\sc geant3}-based simulation~\cite{geant} of detector 
response and processed with the same reconstruction program as used for data. 
Events from randomly selected beam crossings are overlaid on simulated 
events to account for detector noise and 
contributions from additional \ppb\ interactions. 
Parametrizations of trigger
efficiency are determined using events collected with independent triggers 
based on information from the muon detectors. 
Weight factors compensating for residual differences between data 
and simulation are applied for electron, muon and jet 
identification. Jet energy calibration and resolution are adjusted in
simulated events to match those measured in data.


A preselection that greatly reduces the overwhelming background from 
multijet events is performed as follows.
The primary vertex must be reconstructed within the acceptance of the 
silicon vertex detector, 
and at least three tracks must originate from that vertex. 
Jets with associated tracks (using only tracks 
that meet minimal quality criteria to ensure that the 
$b$-tagging algorithm operates efficiently) are denoted as ``taggable'' jets.
There must be two or three taggable jets, one of which 
is the leading (highest \pt) jet. 
These jets must have transverse momentum $\pt>20$~GeV and 
pseudorapidity $\vert\eta\vert<2.5$~\cite{eta}.
The two leading taggable jets must not be back-to-back in the plane 
transverse to the beam direction: $\Delta\phi\jj<165^\circ$. 
Finally,~~$\met>20$~GeV is required. 

Additional selection criteria define four distinct samples: 
(i) an analysis sample used to search for a Higgs boson signal, 
(ii) an electroweak (EW) control sample, 
enriched in $W(\to\mu\nu)$+jets events where the 
jet system has a topology similar to that of the analysis sample, that is
used to validate the SM background simulation, 
(iii) a ``MJ-model'' sample, dominated by multijet events, used to 
model the MJ background in the analysis sample, and
(iv) a large ``MJ-enriched'' sample, used to validate this modeling 
procedure.  

The analysis sample is selected by requiring $\met > 40$~GeV and 
a measure of the \met\ significance ${\cal S} > 5$~\cite{metsig}. 
Larger values of $\cal S$ correspond to \met\ values 
that are less likely to be caused by fluctuations in jet 
energies. In signal events, the missing track \pt, \mpt,
defined as the opposite of the vectorial sum of the charged particle transverse
momenta, is expected to point in a direction close to that of \met.
Such a strong correlation
is not expected in multijet events, where \met\ originates mainly 
from mismeasurement of jet energies. Advantage is taken of this feature by 
requiring ${\cal D}<\pi/2$, where ${\cal D}=\Delta\phi(\met,\mpt)$.
Events containing an isolated electron or muon~\cite{dzerotop}
with $\pt > 15$~GeV are rejected
to reduce backgrounds from $W$+jets, top quark, and diboson production. 

The EW-control sample is selected 
in a way similar to the analysis sample, except that 
an isolated muon with $\pt>15$~GeV is required. 
The multijet content of this sample is rendered negligible by requiring 
the transverse mass of the muon and \met\ system to be larger than 30~GeV.
To  ensure similar jet topologies for the analysis and 
EW-control samples, \met\ not corrected for the selected muon 
is required 
to exceed 40~GeV. Excellent agreement with the SM expectation is found for the 
number of selected events. The agreement for all kinematic distributions 
is also very good once a reweighting of the distribution of $\Delta\eta$ 
between the two leading taggable jets is performed, as suggested by 
a simulation of $(W/Z)$+jets using the {\sc sherpa} generator~\cite{sherpa}.

The MJ-model sample, used to determine the MJ background,
is selected as the analysis sample, 
except that the requirement of 
${\cal D}<\pi/2$ is inverted. The small contribution from non-MJ SM processes 
in the ${\cal D}>\pi/2$ region is subtracted, and the resulting sample 
is used to model the MJ background in the analysis sample. 
After adding contributions from SM backgrounds, the MJ background 
is normalized so that 
the expected number of events is identical to the number observed in the 
analysis sample.
 
The MJ-enriched sample is used to test the validity of this approach and 
is defined as the analysis sample, except that the ~\met\ 
threshold is reduced to 30~GeV and no requirement is 
imposed on $\cal S$. As a result, the MJ background dominates the 
entire range of $\cal D$ values, and this sample is used to verify that 
the events with ${\cal D}>\pi/2$ correctly model those with ${\cal D}<\pi/2$. 

The large branching fraction for $H\to\bbb$ is exploited by 
requiring that one or both of the two 
leading taggable jets be $b$ tagged. The double-tag sample is selected
with asymmetric requirements 
on the outputs of a $b$-tagging neural network algorithm ~\cite{btag}, 
such that one jet is tagged with an 
efficiency of $\approx 70$\% (``loose tag''), 
and the other with an efficiency of $\approx 50$\% (``tight tag''). 
These values apply for taggable jets with 
$\pt\approx 45$~GeV and $\vert\eta\vert\approx 0.8$. 
The mistag rates , i.e., the probablilities to tag light
($u,d,s,g$) jets as $b$ jets, are $\approx 6.5$\% and $\approx 0.5$\%
for the loose and tight tags, respectively.
The sensitivity of the search is improved by defining an independent 
single-tag sample in which one of the two leading taggable 
jets passes the tight tag and the other one fails the loose tag. 
The flavor-dependent $b$-tagging efficiencies are adjusted in simulated events
to match those measured in dedicated data samples.

\begin{figure*}[htbp]
\centering
\begin{tabular}{ccc}
\includegraphics[width=5.5cm]{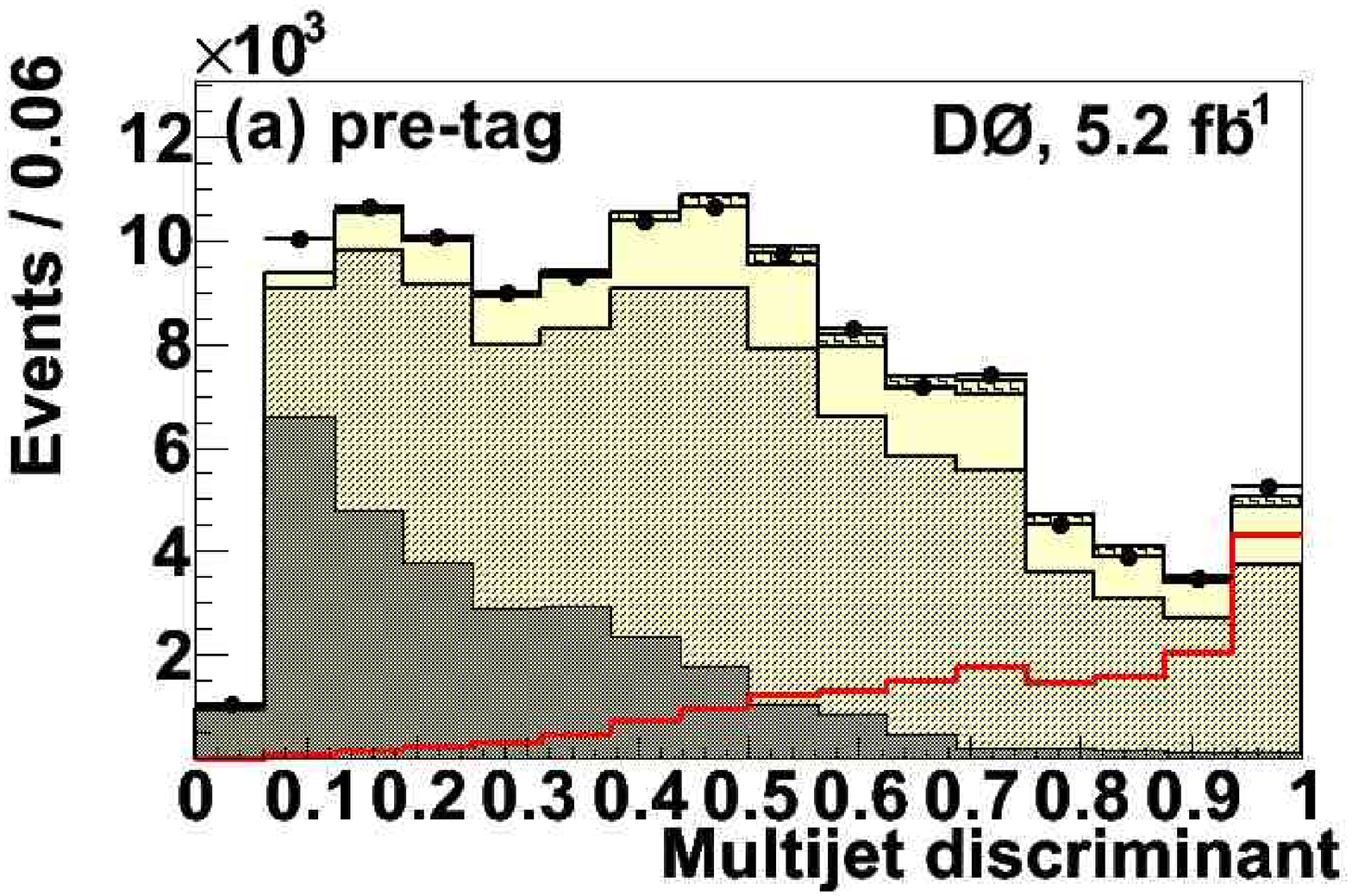} &
\includegraphics[width=5.5cm]{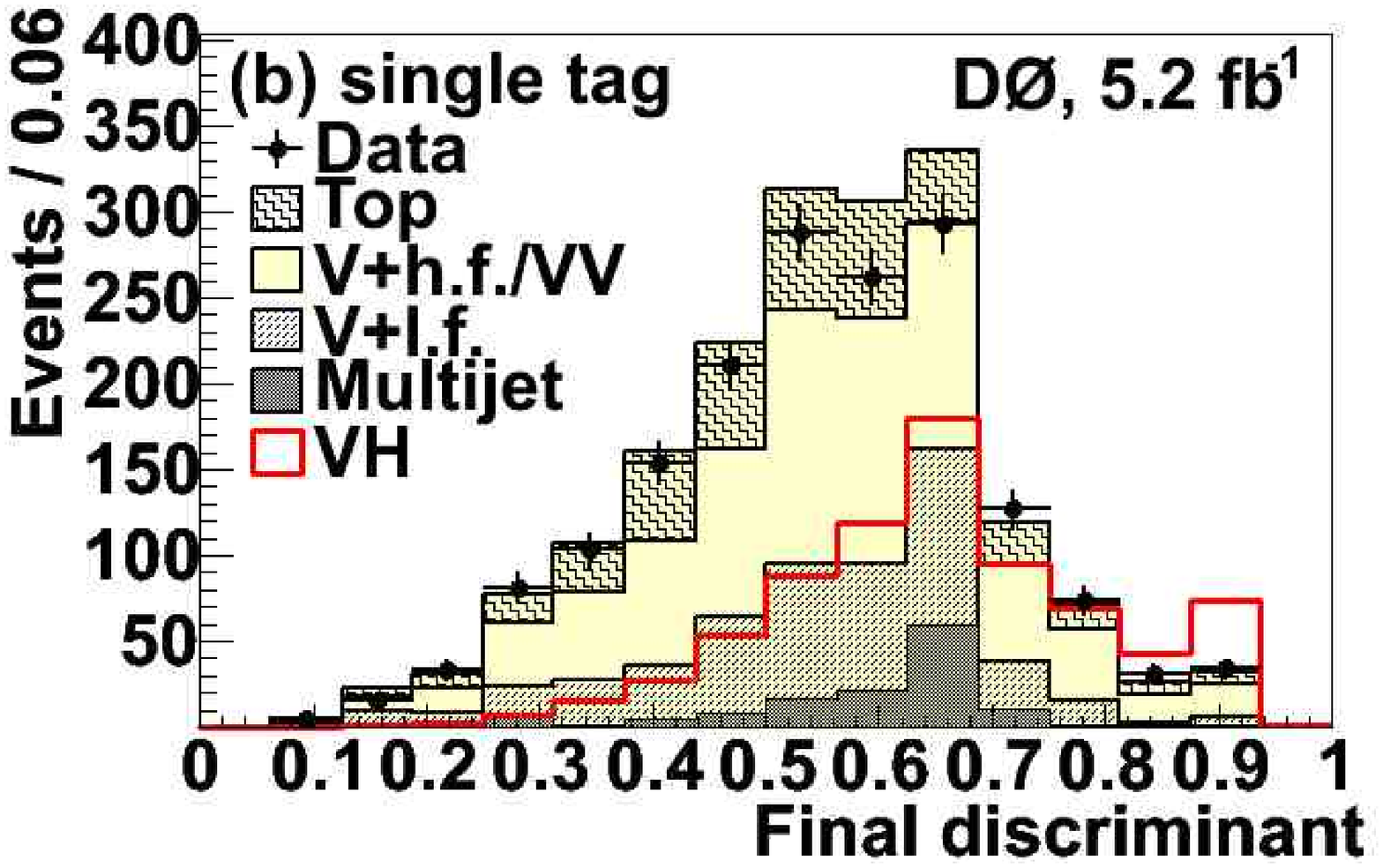} &
\includegraphics[width=5.5cm]{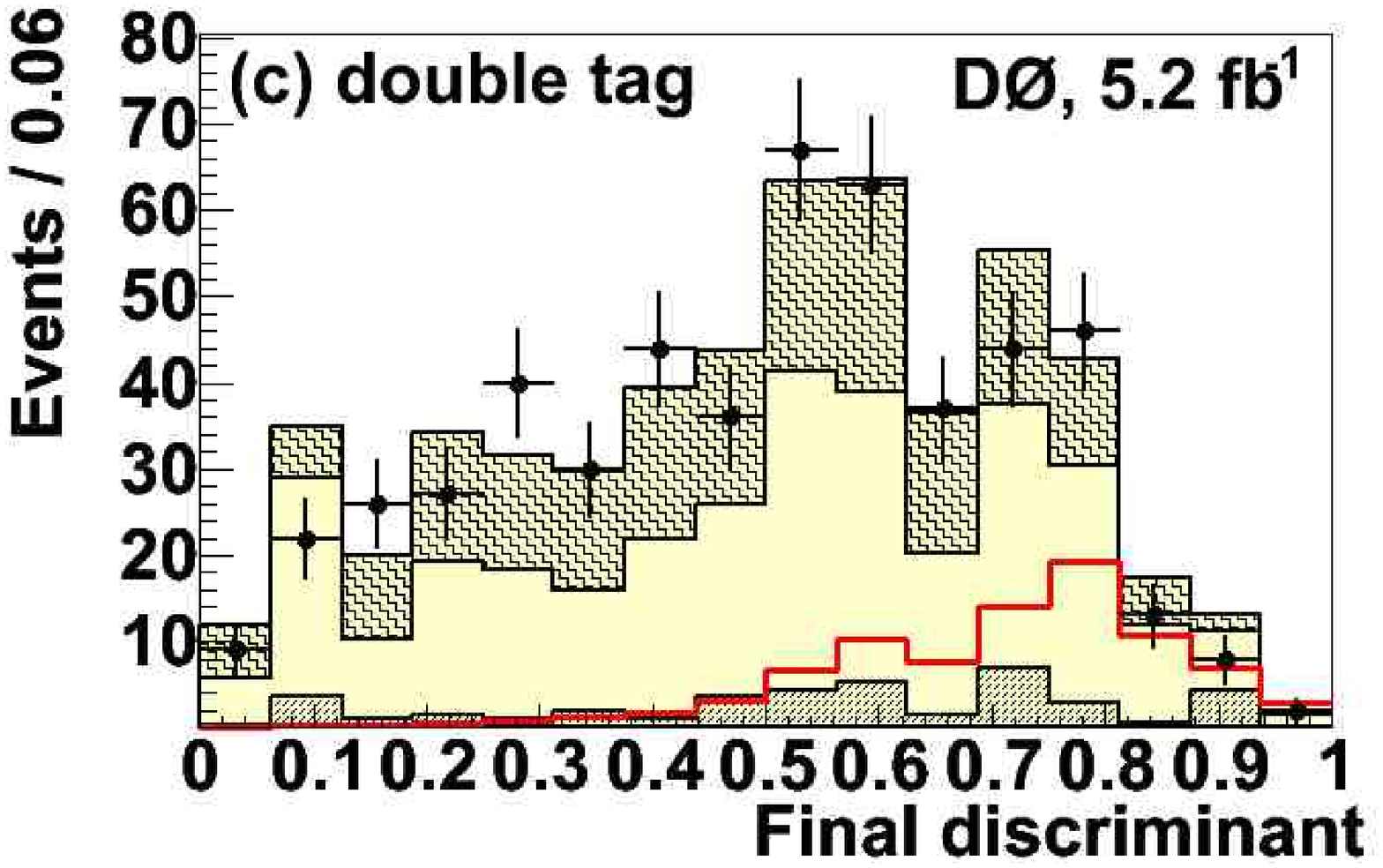}
\end{tabular}
\caption{\label{decision}
Decision tree outputs for $m_H=115$~GeV: (a) for the MJ DT, 
and for the SM DTs following the multijet veto
for (b) single and (c) double tag.
The data are shown as points with error bars. The background
contributions are shown as histograms, with codes 
indicated in the legend in (b). Dibosons are labeled ``VV,''
``V+l.f.'' includes $(W/Z)$+$(u,d,s,g)$ jets, 
``V+h.f.'' includes $(W/Z)$+$(b,c)$ jets, and 
``Top'' includes pair and single top quark production.  
The distributions for signal (VH) are multiplied by 
factors of 500, 100, and 10 in (a)--(c), respectively.}
\end{figure*}

 
A boosted-decision-tree (DT) 
technique~\cite{tree} takes advantage of different 
kinematics in
signal and background processes. 
For each $m_H$, a ``MJ DT'' (multijet-rejection DT),
used to discriminate between signal and
MJ-model events, is trained before $b$ tagging is applied, using 
23 kinematic variables. 
These include the number of jets, jet \pt, dijet \pt, \met, angles between 
jets, between dijet and \met\ and between jets and \met, number of isolated 
tracks, and dijet mass, where the dijet system is constructed from 
the two leading taggable jets. 
The MJ-DT output (multijet discriminant)
is shown in Fig.~\ref{decision}(a) for $m_H=115$~GeV. 
A value of the multijet discriminant
in excess of 0.6 is required (multijet veto), which removes over 95\%
of the multijet background and 65\% of the non-MJ SM backgrounds, while 
retaining 70\% of the signal. The number of expected signal and 
background events, as well as the number of observed events, 
are given in Table~\ref{yields} after imposing the multijet veto. 

\begin{table*}
\caption{\label{yields} The number of expected signal 
and background events, and the number observed after the 
multijet veto, prior to $b$ tagging and for
single and double tags. The
signal corresponds to $m_H=115$~GeV, ``Top'' includes pair and 
single top quark 
production, and $VV$ is the sum of all diboson processes. 
The uncertainties quoted arise from the statistics of the simulation
and from the sources of systematic uncertainties mentioned in the text.}
\begin{ruledtabular}
\begin{tabular}{lccccccccc}
Sample & $ZH$ & $WH$ & $W$+jets & $Z$+jets & Top & $VV$ & Multijet & Total background & Observed \\
\hline
Pretag & 13.73 $\pm$ 1.37 & 11.64 $\pm$ 1.17 & 19\,069 & 9432 & 1216 & 1112 & 
1196 & 32\,025 $\pm$ 4121 & 31\,718 \\
Single tag & ~\,4.16 $\pm$ 0.42 & ~\,3.60 $\pm$  0.37 & 802 & 439 & 404 & 60 & 
125 & ~1830 $\pm$ 273 & 1712 \\
Double tag & ~\,4.66 $\pm$ 0.58 & ~\,4.00 $\pm$  0.50 & 191 & 124 & 199 & 24 & 
$<8$ & ~538 $\pm$ 93 & 514 \\
\end{tabular}
\end{ruledtabular}
\end{table*}

To discriminate 
signal from SM backgrounds, additional ``SM DTs'' (SM rejection DTs) are 
trained separately  
for the single and double-tag samples, using the
same kinematic variables as for the MJ DT.
The outputs of the SM DTs after the multijet veto
(final discriminants) are shown in 
Figs.~\ref{decision}(b) and \ref{decision}(c) for $m_H=115$~GeV, for the
single and double tag samples.

Agreement between data and expectation from SM and MJ
backgrounds is observed in the single and double tag samples, 
once the systematic uncertainties discussed below are taken into account,
both in the number of selected events (Table~\ref{yields})
and in distributions of final discriminants (Fig.~\ref{decision}). 
A modified frequentist approach~\cite{cls} is used
to set limits on the cross section for SM Higgs boson production, where
the test statistic is a joint log-likelihood ratio (LLR)
of the background-only and signal+background hypotheses,
obtained by summing LLR values over the bins in 
the final discriminants shown in Figs.~\ref{decision}(b) and \ref{decision}(c).
The impact of systematic uncertainties on the sensitivity 
of the analysis is reduced by maximizing a ``profile''
likelihood function~\cite{wade} in which these uncertainties 
are given Gaussian constraints associated with their priors.


Experimental uncertainties arise from trigger simulation (3\%), 
jet energy calibration and resolution (3\% for signal and $4\%-5\%$ 
for background), 
jet reconstruction and taggability ($2\%-3\%$), 
lepton identification ($1\%-2\%$), 
and $b$ tagging (from 2\% for signal in the single-tag sample 
to 8\% for background in the double-tag sample). 
Their impact is assessed on overall
normalizations and shapes of distributions in final discriminants. 
Correlations among systematic uncertainties in signal and 
background are taken into account in extracting the final results, 
including a 6.1\% uncertainty on the integrated luminosity.

Theoretical uncertainties on cross sections for SM processes are 
estimated as follows.
For $(W/Z)$+jets production, an uncertainty of 6\% 
is assigned to the total cross sections, and an uncertainty 
of 20\% on the heavy-flavor fractions (estimated from {\sc mcfm}). 
For other SM backgrounds, uncertainties are taken from~\cite{xsections} 
or from {\sc mcfm}, and range from 6\% to 10\%. 
The uncertainties on cross sections for signal (6\% for $m_H=115$~GeV) 
are taken from~\cite{signal}. Uncertainties on the shapes of the final 
discriminants arise from (i) the modeling of
$(W/Z)$+jets, assessed by varying the 
renormalization-and-factorization scale and by comparing {\sc alpgen} 
interfaced with {\sc herwig}~\cite{herwig} to {\sc alpgen}
interfaced with {\sc pythia}, and (ii) the
choice of PDFs, estimated using the prescription of~\cite{cteq}.
The normalization of the MJ background 
is anticorrelated with the normalization of the SM 
backgrounds, as the sum is constrained by data prior to $b$ tagging.

\begin{figure}[htbp]
\centering
\includegraphics[width=8.5cm]{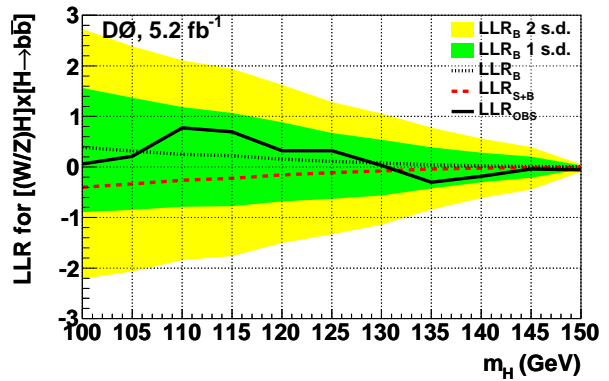}
\caption{\label{final} 
The observed LLR is shown as a solid black line, the expected 
LLRs for the background-only and 
signal+background hypotheses are shown as black dots and red dashes, 
respectively, 
and the heavy green and light yellow shaded areas 
correspond to 1 and 2 standard deviations (s.d.) around the expected 
LLR for the background-only hypothesis.}
\end{figure}

The results of the analysis are given as limits in Table~\ref{limits} and as 
LLRs in Fig.~\ref{final}, as a function of $m_H$.
The observed LLRs are within 1 standard deviation of expectation 
(the median of the LLR for the background-only hypothesis).
For $m_H=115$~GeV, the observed and expected limits
on the combined cross section of $ZH$ and $WH$ production, multiplied by
the branching 
fraction for $H\to\bbb$, are factors of 3.7 and 4.6 larger than the 
SM value, respectively.
These are the most constraining results for a SM Higgs boson decaying 
dominantly into $\bbb$ for $m_H$ above the limit set at LEP. 

\begin{table*}
\caption{\label{limits} 
As a function of $m_H$, observed and expected upper limits on the $(W/Z)H$ 
production cross section multiplied by branching fraction for $H\to\bbb$, 
relative to the SM expectation. 
}
\begin{ruledtabular}
\begin{tabular}{lccccccccccc}
$m_H$ (GeV) & 100 & 105 & 110 & 115 & 120 & 125 & 130  & 135  & 140  & 145  & 150  \\
\hline
Observed & 3.6 & 3.9 & 3.4 & 3.7 & 4.9 & 5.5 & 7.4 & 14.2 & 18.0 & 20.9 & 37.5 \\
Expected & 3.4 & 3.8 & 4.2 & 4.6 & 5.5 & 6.7 & 7.8 & 10.5 & 14.7 & 21.2 & 33.6 \\
\end{tabular}
\end{ruledtabular}
\end{table*}

Supplementary material is provided in~\cite{EPAPS}.


\begin{acknowledgments}
%
We thank the staffs at Fermilab and collaborating institutions, 
and acknowledge support from the 
DOE and NSF (USA);
CEA and CNRS/IN2P3 (France);
FASI, Rosatom and RFBR (Russia);
CNPq, FAPERJ, FAPESP and FUNDUNESP (Brazil);
DAE and DST (India);
Colciencias (Colombia);
CONACyT (Mexico);
KRF and KOSEF (Korea);
CONICET and UBACyT (Argentina);
FOM (The Netherlands);
STFC and the Royal Society (United Kingdom);
MSMT and GACR (Czech Republic);
CRC Program, CFI, NSERC and WestGrid Project (Canada);
BMBF and DFG (Germany);
SFI (Ireland);
The Swedish Research Council (Sweden);
and
CAS and CNSF (China).
%
\end{acknowledgments}

\begin{table*}
\begin{center}
{\Large{\bf Supplementary material}}
\end{center}
\end{table*}

\begin{table*}
\caption{\label{tbl:ZsecBR}
Theoretical cross sections for associated $WH$ and $ZH$ production and 
$H\to\bbb$ branching fraction, as a function of $m_H$.
}
 \begin{center}
  \begin{tabular}{lccccccccccc}
\hline
\hline
$m_H$ (GeV) & 100 & 105 & 110 & 115 & 120 & 125 & 130  & 135  & 140  & 145  & 150  \\
$\sigma(WH)$ (pb) & 0.286 & 0.243 & 0.208 & 0.178 & 0.153 & 0.132 & 0.115 & 0.099 & 0.087 & 0.076 & 0.066 \\
$\sigma(ZH)$ (pb) & 0.167 & 0.143 & 0.123 & 0.107 & 0.093 & 0.081 & 0.070 & 0.062 & 0.054 & 0.048 & 0.042 \\
$B(H\to\bbb)$ & 0.812 & 0.796 & 0.770 & 0.732 & 0.679 & 0.610 & 0.527 & 0.436 & 0.344 & 0.256 & 0.176 \\
\hline
\hline
  \end{tabular}
 \end{center}
\end{table*}

\begin{table*}
\caption{\label{tbl:p17p20selCuts}
The number of observed events and the number of $ZH$ and $WH$ signal events 
expected for $m_H=115$~GeV at different stages of the selection.
}
 \begin{center}
  \begin{tabular}{lccc}
\hline
\hline
                                 & Data & $ZH$ & $WH$ \\
\hline
Preselection                     & 7690773  & 19.9 & 40.8 \\
$\met > 40$~GeV                  &  790496  & 19.2 & 36.1 \\
\met\ Significance $> 5$         &  188761  & 18.2 & 32.7 \\
Isolated $e/\mu$ veto            &  153542  & 18.1 & 21.4 \\
$\Delta\phi(\met,\mpt)<\pi/2$    &  120875  & 17.7 & 18.6 \\
\hline
\hline
  \end{tabular}
 \end{center}
\end{table*}

\begin{table*}
\caption{\label{tbl:eventCounts}
The number of expected signal and background events, and the number observed in the analysis sample before the multijet veto, prior to $b$ tagging and for single and double tags; ``top'' includes pair and single top quark production. The quoted uncertainties are statistical only.}
 \begin{center}
  \begin{tabular}{l|rl|rl|rl}
\hline
\hline
   Sample         &\multicolumn{2}{c|}{pre-tag} &\multicolumn{2}{c|}{single tag} &\multicolumn{2}{c}{double tag}\\
   \hline
   $ZH$ (115 \Gev) &    17.72 & $\pm$   0.09 &    5.44 & $\pm$   0.05 &    5.69 & $\pm$ 0.05 \\
   $WH$ (115 \Gev) &    18.55 & $\pm$   0.15 &    5.81 & $\pm$   0.08 &    5.83 & $\pm$ 0.07 \\
   \hline
   $W+$jets        & 55502 & $\pm$ 135 & 1311 & $\pm$  24 &  136 & $\pm$ 10 \\
   $W+$b/c jets    &  9102 & $\pm$  46 & 1252 & $\pm$  15 &  411 & $\pm$ 8 \\
   $Z+$jets        & 17785 & $\pm$ 131 &  211 & $\pm$  17 &    9 & $\pm$ 3 \\
   $Z+$b/c jets    &  4621 & $\pm$  36 &  701 & $\pm$  11 &  256 & $\pm$ 6 \\
   top             &  2408 & $\pm$   6 &  815 & $\pm$   3 &  427 & $\pm$ 2 \\
   di-boson        &  2309 & $\pm$  15 &  126 & $\pm$   3 &   42 & $\pm$ 2 \\
   \hline
   SM background         & 91727 & $\pm$ 197 & 4415 & $\pm$  35 & 1282 & $\pm$ 15 \\
   MJ background         & 29148 & $\pm$ 377 & 2255 & $\pm$ 101 &  398 & $\pm$ 20 \\
   \hline
   Total background      &120875 & $\pm$ 425 & 6670 & $\pm$ 107 & 1679 & $\pm$ 25 \\
   Observed        &\multicolumn{2}{c|}{120875} & \multicolumn{2}{c|}{6853} & \multicolumn{2}{c}{1581}\\
\hline
\hline
  \end{tabular}
 \end{center}
\end{table*}

\begin{table*}
\caption{\label{tbl:eventCountsMjdt}
The number of expected signal and background events, and the number observed in the analysis sample after the multijet veto, prior to $b$ tagging and for single and double tags; ``top'' includes pair and single top quark production. The quoted uncertainties are statistical only.}
 \begin{center}
  \begin{tabular}{l|rl|rl|rl}
\hline
\hline
   Sample          &\multicolumn{2}{c|}{pre-tag} &\multicolumn{2}{c|}{single tag} &\multicolumn{2}{c}{double tag}\\
   \hline
   $ZH$ (115 \Gev) &    13.73 & $\pm$   0.08 &    4.16 & $\pm$  0.05 &   4.66 & $\pm$ 0.04 \\
   $WH$ (115 \Gev) &    11.64 & $\pm$   0.12 &    3.60 & $\pm$  0.07 &   3.99 & $\pm$ 0.06 \\
   \hline
   $W+$jets        & 15997 & $\pm$  65 &  367 & $\pm$ 12 &  38 & $\pm$ 6 \\
   $W+$b/c jets    &  3072 & $\pm$  26 &  435 & $\pm$  8 & 153 & $\pm$ 5 \\
   $Z+$jets        &  7304 & $\pm$  80 &   94 & $\pm$ 12 &   2 & $\pm$ 1 \\
   $Z+$b/c jets    &  2129 & $\pm$  24 &  344 & $\pm$  8 & 122 & $\pm$ 4 \\
   top             &  1216 & $\pm$   4 &  404 & $\pm$  2 & 199 & $\pm$ 2 \\
   di-boson        &  1112 & $\pm$  10 &   60 & $\pm$  2 &  24 & $\pm$ 1 \\
   \hline
   SM background         & 30829 & $\pm$ 109 & 1704 & $\pm$ 20 & 539 & $\pm$ 9 \\
   MJ background         &  1196 & $\pm$ 120 &  125 & $\pm$ 32 &  -1 & $\pm$ 8 \\
   \hline
   Total background      & 32025 & $\pm$ 162 & 1830 & $\pm$ 38 & 538 & $\pm$ 12 \\
   Observed        &\multicolumn{2}{c|}{31718} & \multicolumn{2}{c|}{1712} & \multicolumn{2}{c}{514} \\
\hline
\hline
  \end{tabular}
 \end{center}
\end{table*}

\begin{table*}
\caption{\label{tbl:DTvarsQCD} Variables used as input to the Decision Trees.} 
\begin{center}
\begin{tabular}{l}
\hline
\hline
Number of jets\\	
Number of taggable jets\\	
leading jet $p_{T}$\\
second jet $p_{T}$\\
third jet $p_{T}$\\	
\hht (scalar sum of jet \pt)\\
$\Delta R(\mathrm{jet}_1,\mathrm{jet}_2)$\\
$\Delta\eta(\mathrm{jet}_1,\mathrm{jet}_2)$\\
$\Delta\phi(\mathrm{jet}_1,\mathrm{jet}_2)$\\
\met \\
\met\ significance\\
$\Delta\phi(\met ,\mathrm{jet}_1)$\\
$\Delta\phi(\met ,\mathrm{jet}_2)$\\
$\Delta\phi(\met ,\mathrm{dijet~system})$\\
$\min\Delta\phi(\met,\mathrm{jet_i})$\\
$\max\Delta\phi(\met,\mathrm{jet_i})+\min\Delta\phi(\met,\mathrm{jet_i})$\\
$\max\Delta\phi(\met,\mathrm{jet_i})-\min\Delta\phi(\met,\mathrm{jet_i})$\\
\mht (vectorial sum of jet \pt)\\
\mht / \hht \\
dijet \pt \\
dijet mass\\
dijet transverse mass\\
Number of isolated tracks\\
\hline
\hline
\end{tabular}
\end{center}
\end{table*}

\begin{table*}
\caption{\label{tab:systematics}
Systematic uncertainties in \% of the overall signal and background yields. 
``Jet EC'' and ``Jet ER'' stand for jet 
energy calibration and resolution, respectively. 
``Jet R\&T'' stands for jet reconstruction and taggability.
``Signal'' includes $ZH$ 
and $WH$ production for $m_H=115$~GeV.}
\begin{center}
\begin{tabular}{l|c|c}
\hline
\hline
Systematic Uncertainty & Signal & Background \\
\hline
\multicolumn{3}{c}{pre-tag} \\
\hline
Jet EC -- Jet ER &    2.7 &   7.7 \\ 
Jet R\&T &    3.0 &   3.7 \\ 
Trigger &    2.9 &   3.1 \\ 
Lepton identification &    1.0 &   1.1 \\ 
Heavy-flavor fractions &    -- &   2.6 \\ 
Cross sections &    6.0 &   6.3 \\ 
Luminosity &    6.1 &   5.9 \\ 
Multijet normalization &    -- &   0.9 \\ 
\hline
Total &   10.0 & 12.9 \\ 
\hline
\multicolumn{3}{c}{single tag} \\
\hline
Jet EC -- Jet ER &    2.6 &   4.7 \\ 
Jet R\&T &    3.0 &   2.5 \\ 
$b$ tagging &    1.9 &   5.2 \\ 
Trigger &    2.9 &   3.0 \\ 
Lepton identification &    1.0 &   1.2 \\ 
Heavy-flavor fractions &    -- &   8.1 \\ 
Cross sections &    6.0 &   7.1 \\ 
Luminosity &    6.1 &   5.7 \\ 
Multijet normalization &    -- &   1.8 \\ 
\hline
Total &  10.1 & 14.8 \\ 
\hline
\multicolumn{3}{c}{double tag} \\
\hline
Jet EC -- Jet ER &    2.8 &   3.6 \\ 
Jet R\&T &    3.2 &   2.2 \\ 
$b$ tagging &    7.3 &   8.0 \\ 
Trigger &    3.0 &   3.3 \\ 
Lepton identification &    1.1 &   1.6 \\ 
Heavy-flavor fractions &    -- &   9.8 \\ 
Cross sections &    6.0 &   8.0 \\ 
Luminosity &    6.1 &   6.1 \\ 
Multijet normalization &    -- &   0.4 \\ 
\hline
Total &  12.4 & 17.1 \\ 
\hline
\hline
\end{tabular}
\end{center}
\end{table*}

\begin{figure*}[htp]
\centering
\subfigure[]{\includegraphics[width=8.5cm]{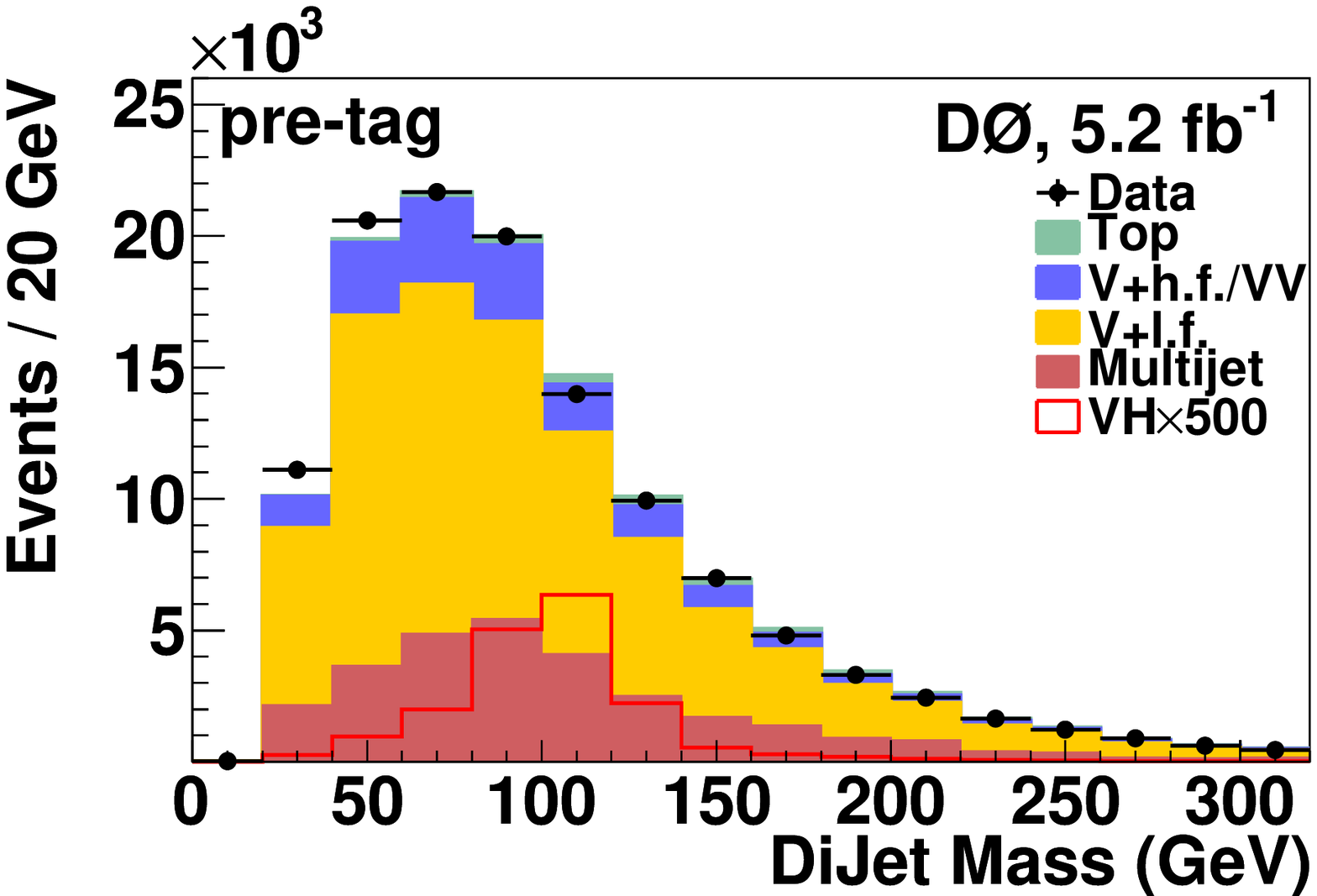}}
\subfigure[]{\includegraphics[width=8.5cm]{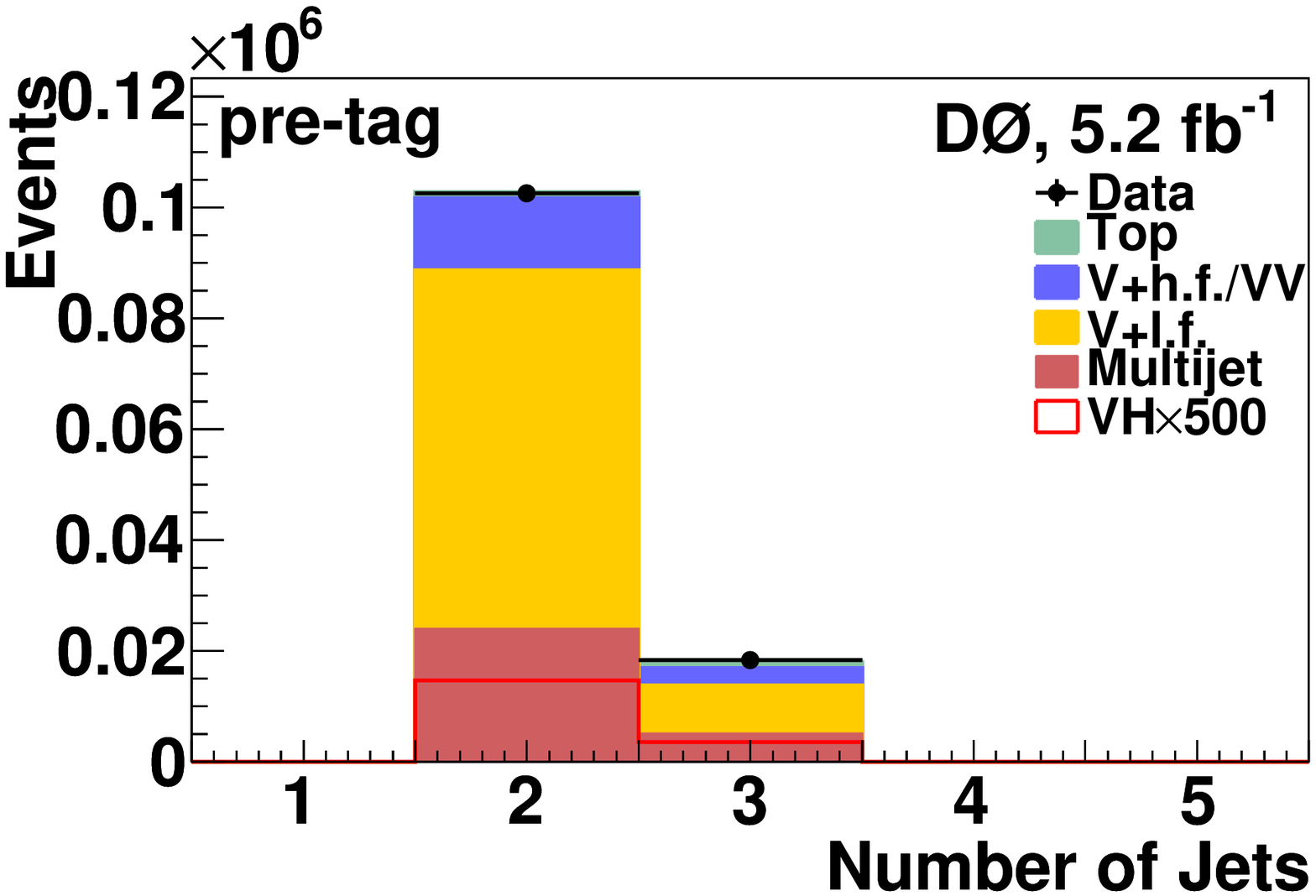}}
\subfigure[]{\includegraphics[width=8.5cm]{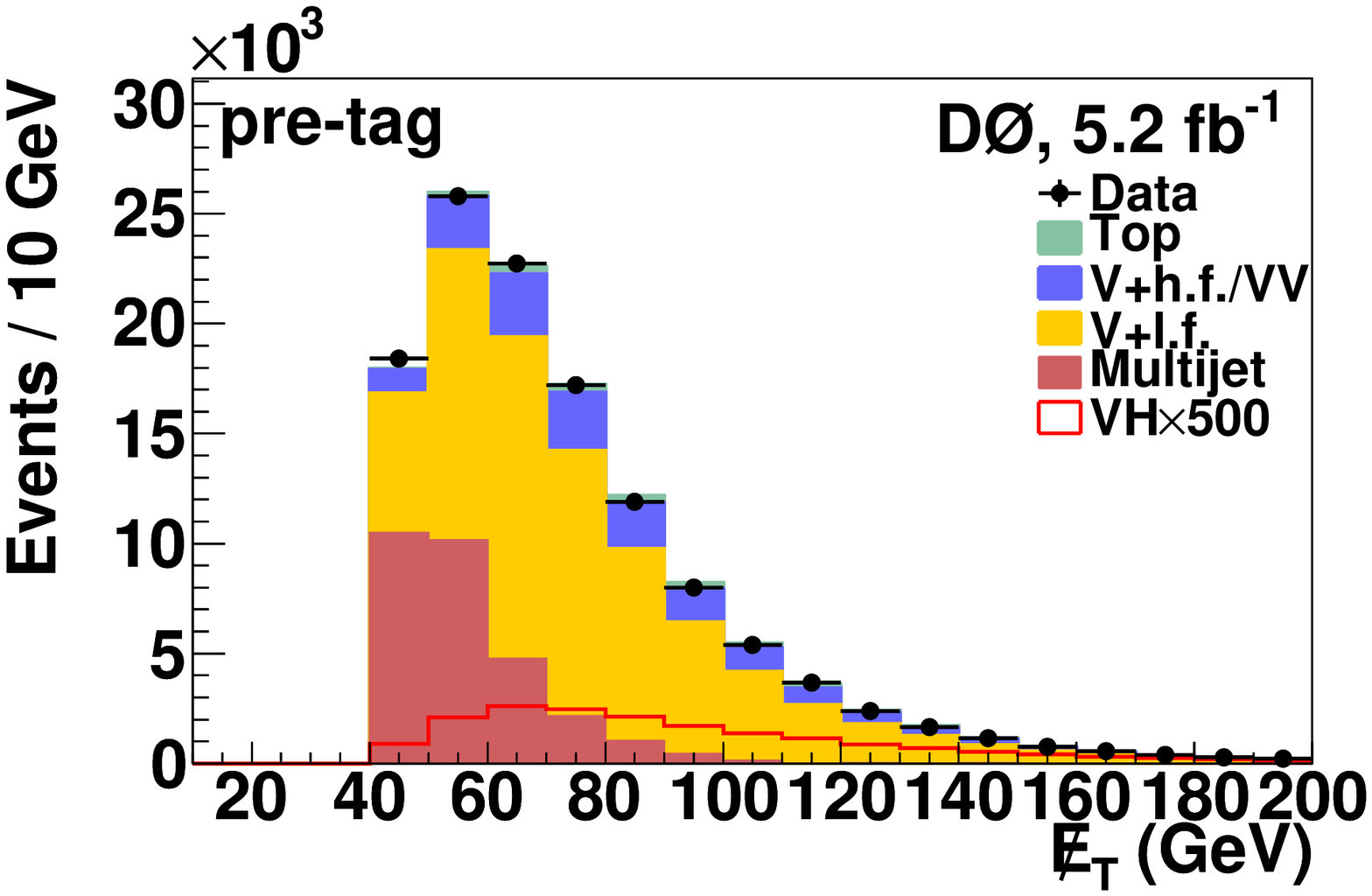}}
\subfigure[]{\includegraphics[width=8.5cm]{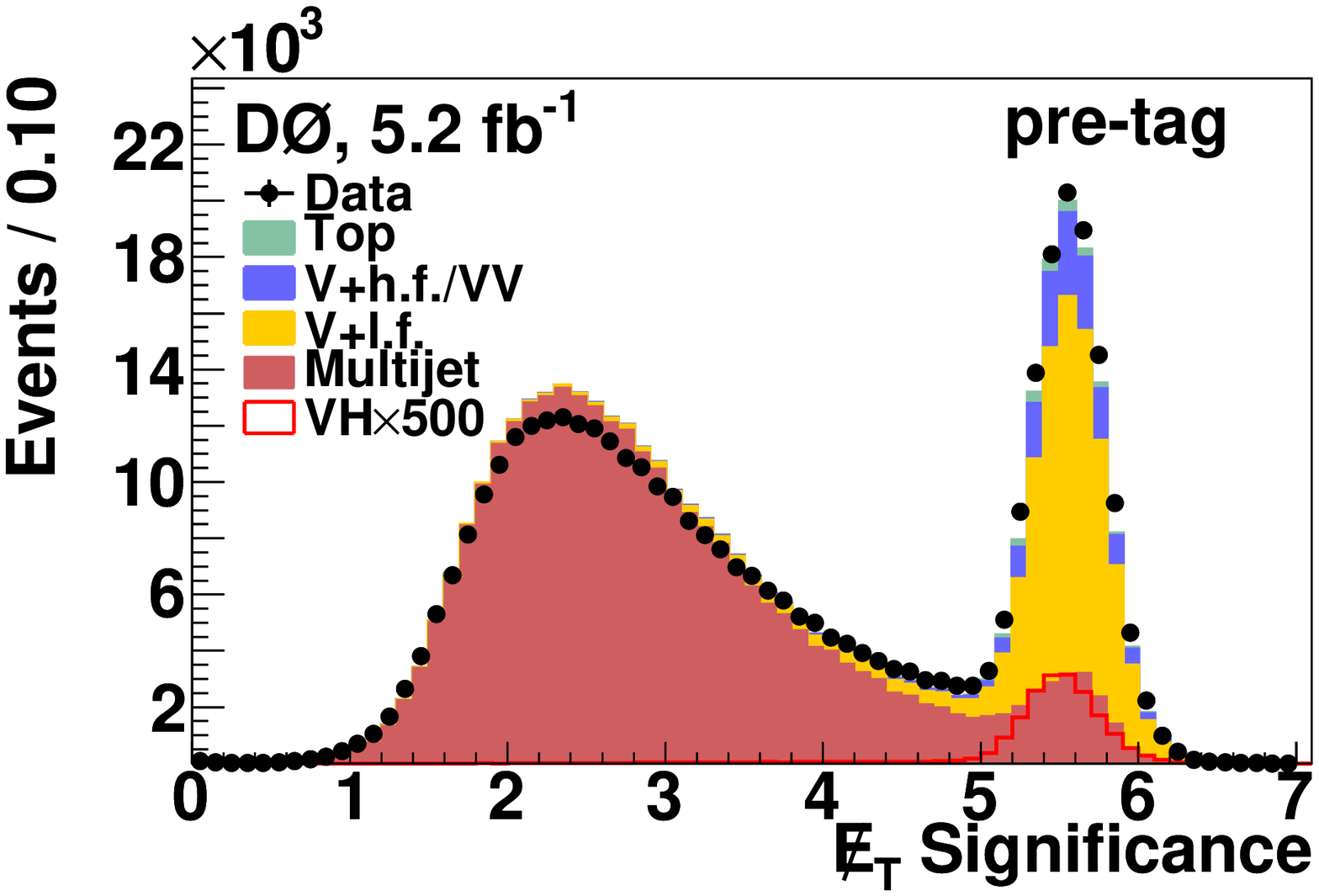}}
\caption{\label{HM_analysis_pretag}  
Distributions in the analysis sample before the multijet veto:
(a) Dijet invariant mass,
(b) Taggable jet multiplicity,
(c) Missing \et,
(d) Missing \et\ significance without the requirement that it be larger than 5.
The signal includes $ZH$ and $WH$ production for $m_H=115$~GeV.
}
\end{figure*}

\begin{figure*}[htp]
\centering
\subfigure[]{\includegraphics[width=8.5cm]{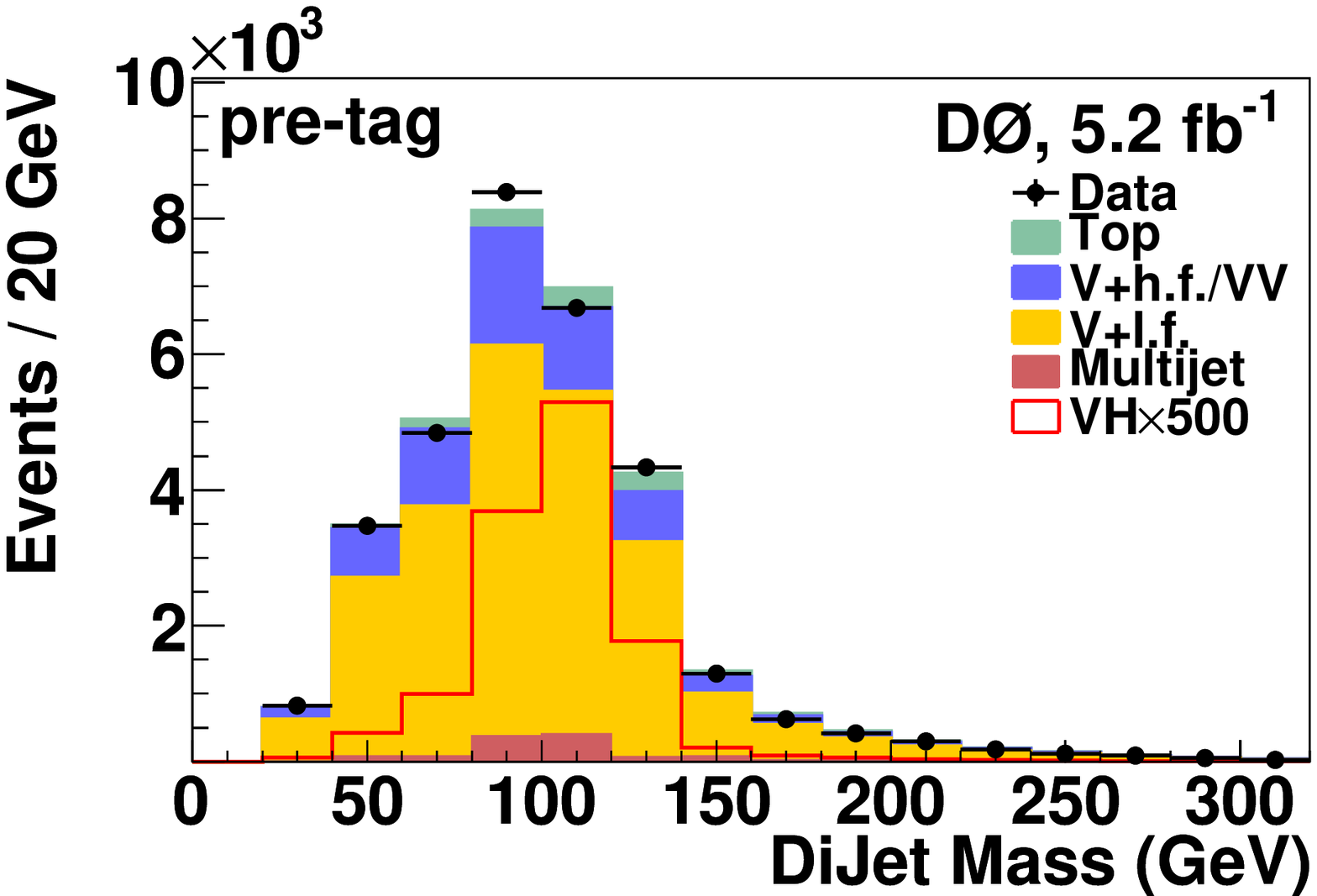}}
\subfigure[]{\includegraphics[width=8.5cm]{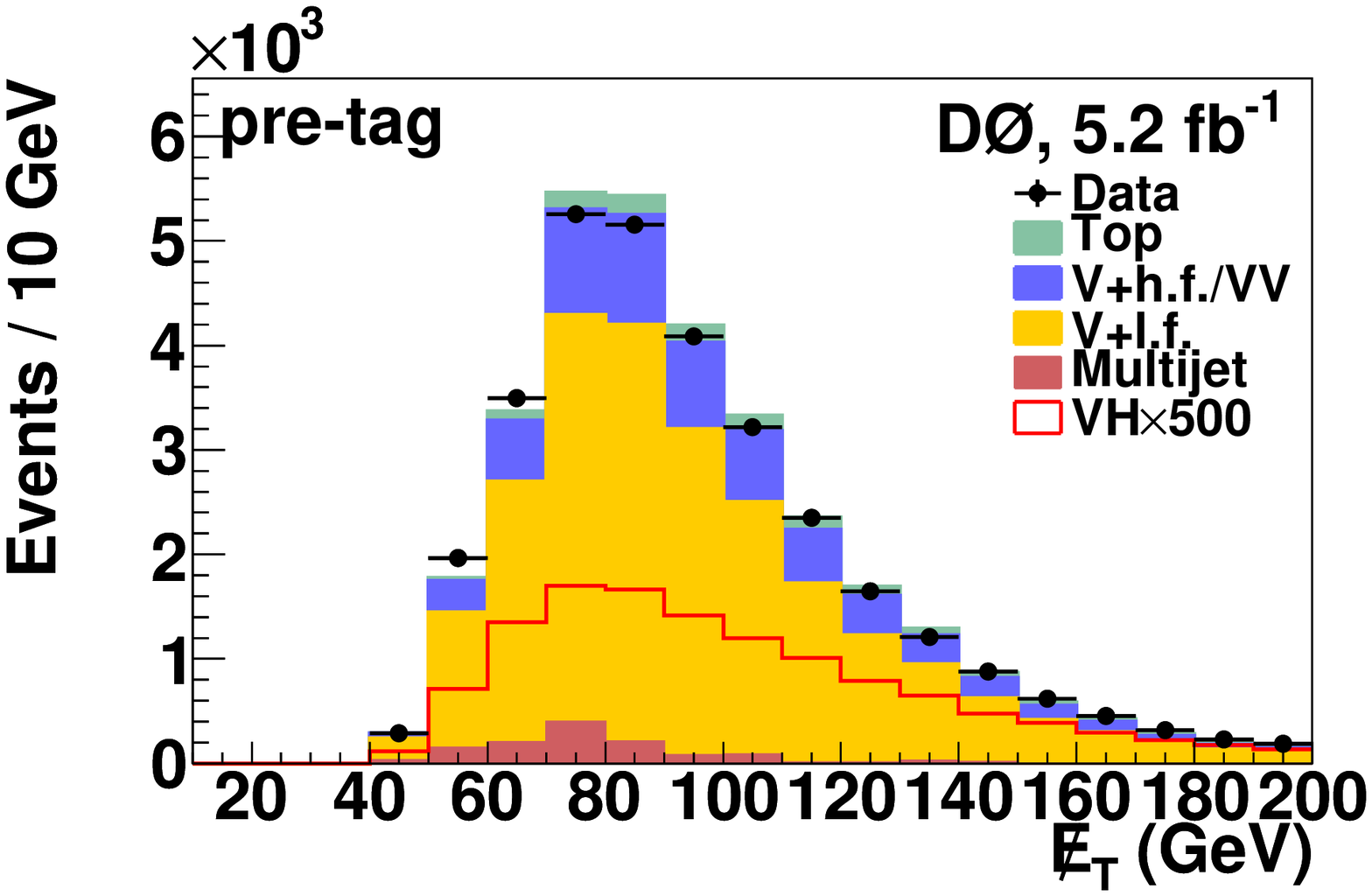}}
\subfigure[]{\includegraphics[width=8.5cm]{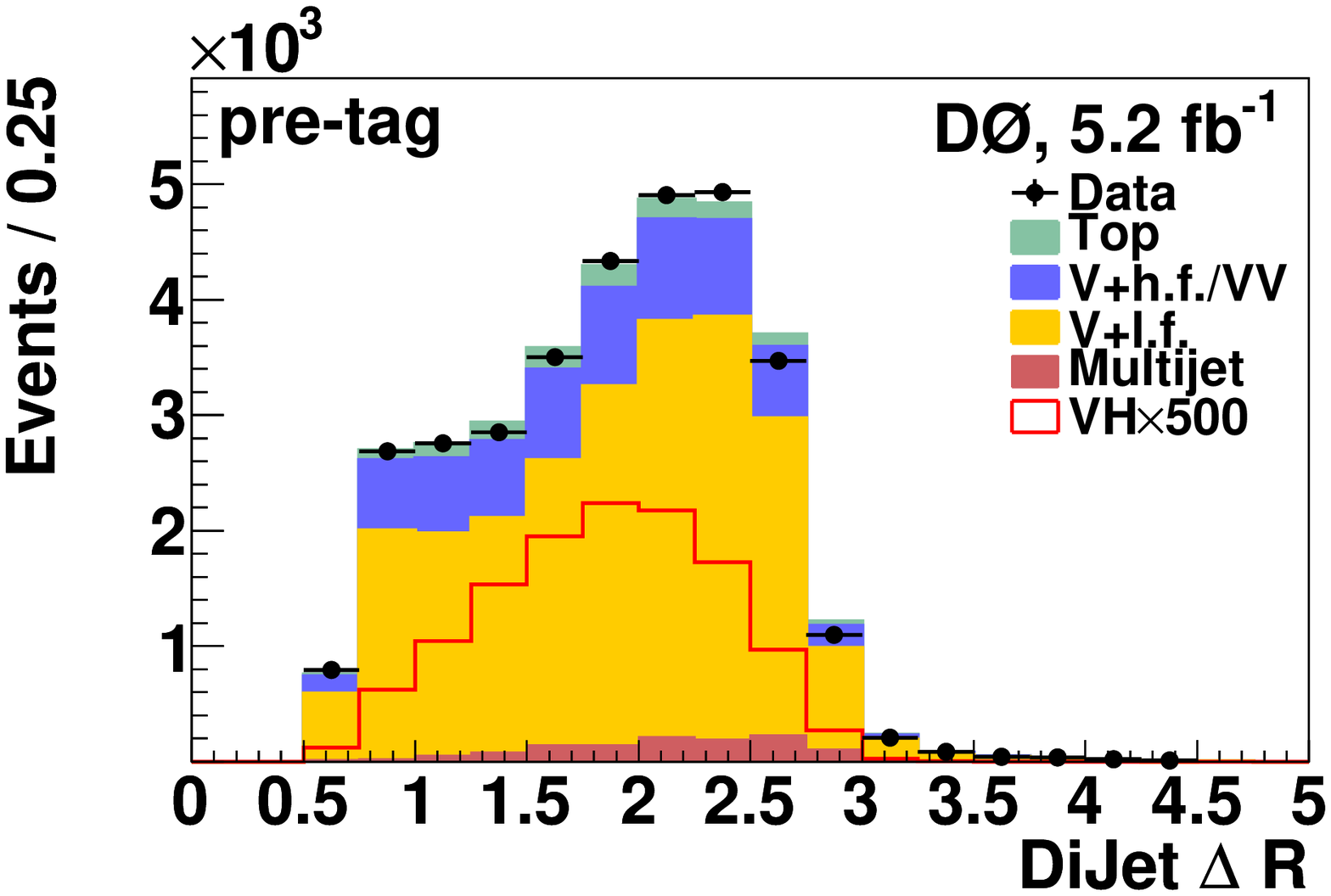}}
\subfigure[]{\includegraphics[width=8.5cm]{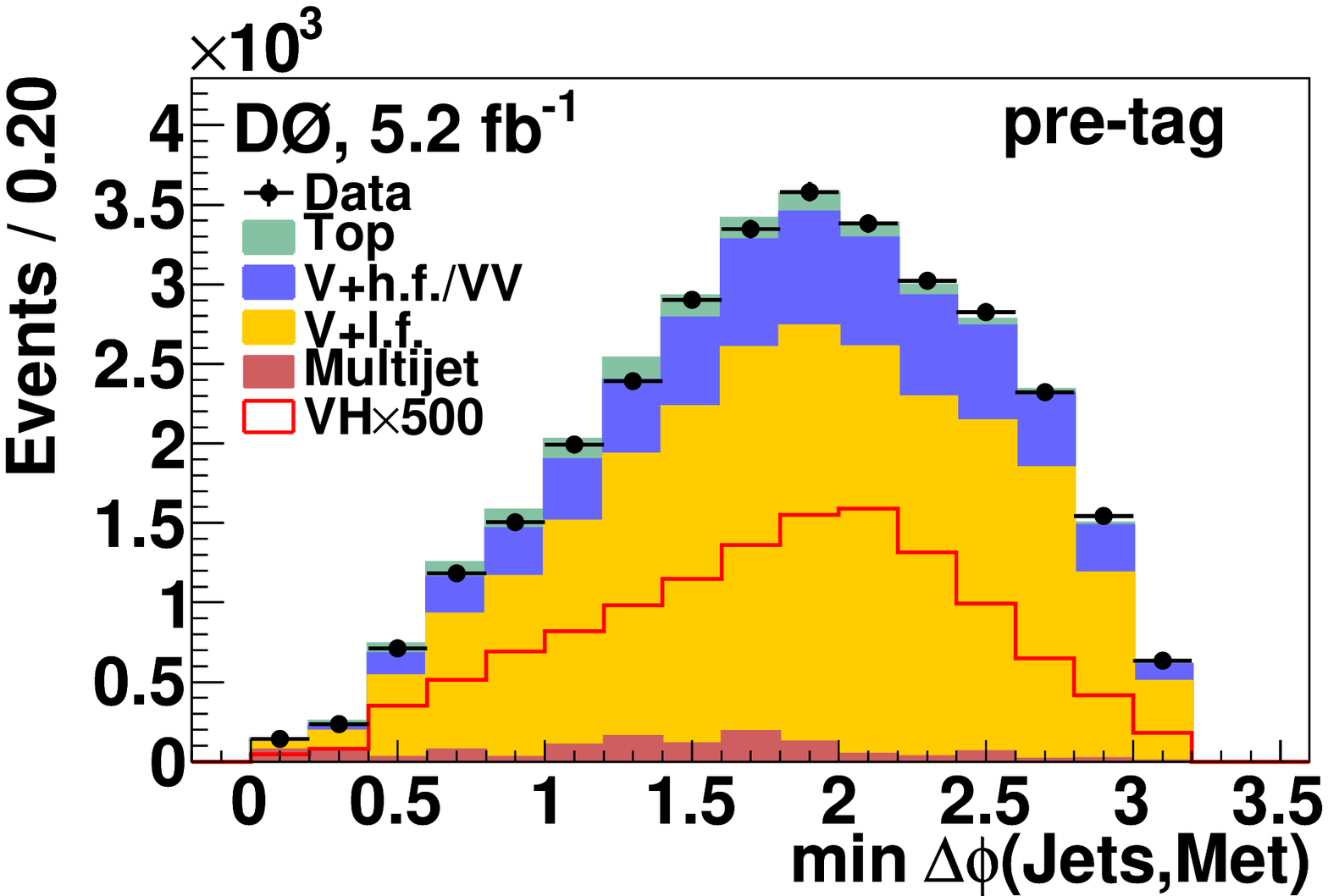}}
\caption{\label{JetDR_analysis_mjveto_pretag}
Distributions in the analysis sample after the multijet veto:  
(a) Dijet invariant mass,
(b) Missing \et,
(c) Dijet $\Delta R$,
(d) Minimum $\Delta\phi$ between any jet and \met.
The signal includes $ZH$ and $WH$ production for $m_H=115$~GeV.
}
\end{figure*}

\begin{figure*}[htp]
\centering
\subfigure[]{\includegraphics[width=8.5cm]{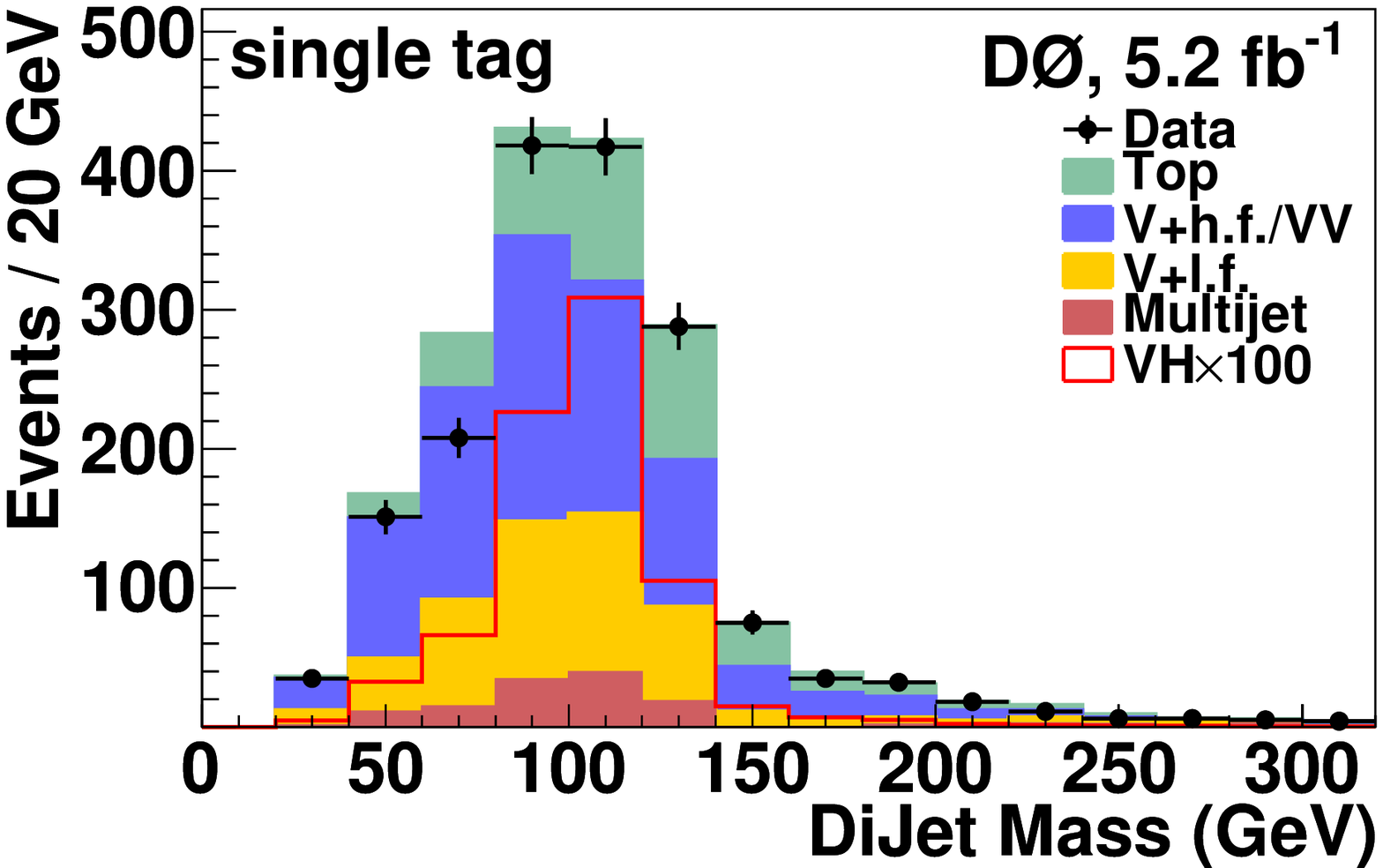}}
\subfigure[]{\includegraphics[width=8.5cm]{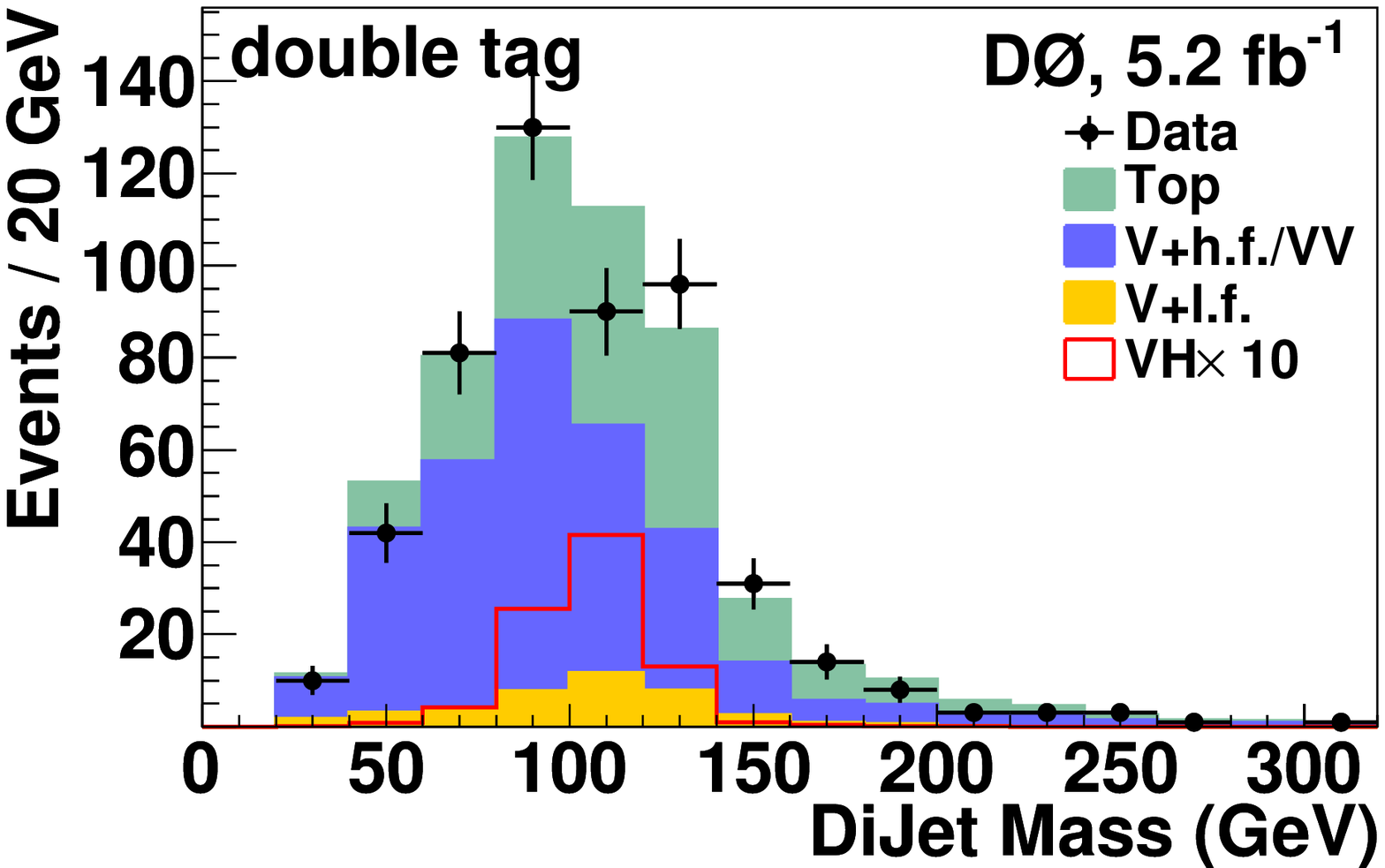}}
\caption{\label{HM_analysis_mjveto_1tag}
Distributions in the analysis sample after the multijet veto:  
(a) Dijet invariant mass with single tag,
(b) Dijet invariant mass with double tag.
The signal includes $ZH$ and $WH$ production for $m_H=115$~GeV.
}
\end{figure*}

\begin{figure*}[htp]
\centering
\subfigure[]{\includegraphics[width=8.5cm]{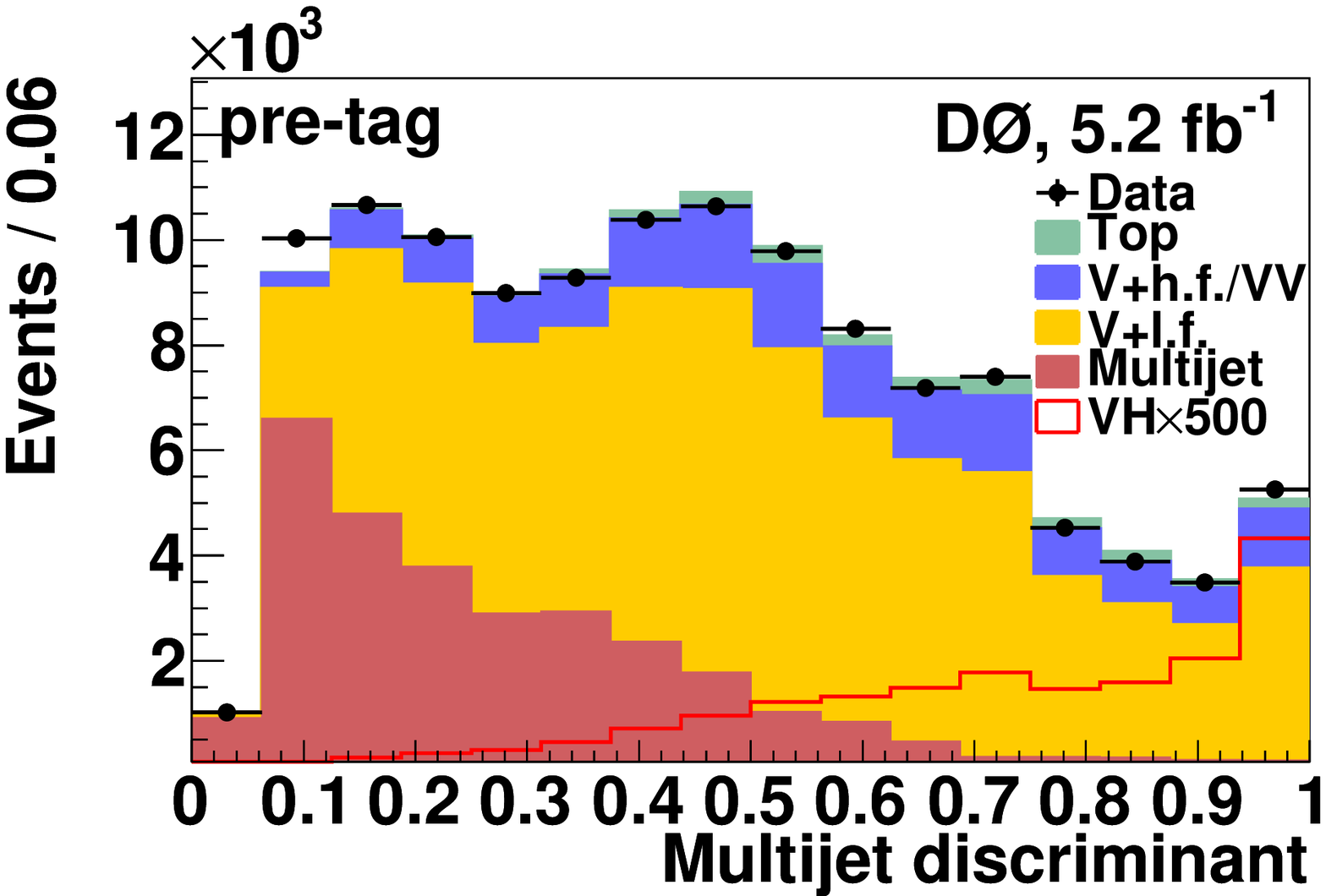}} \\
\subfigure[]{\includegraphics[width=8.5cm]{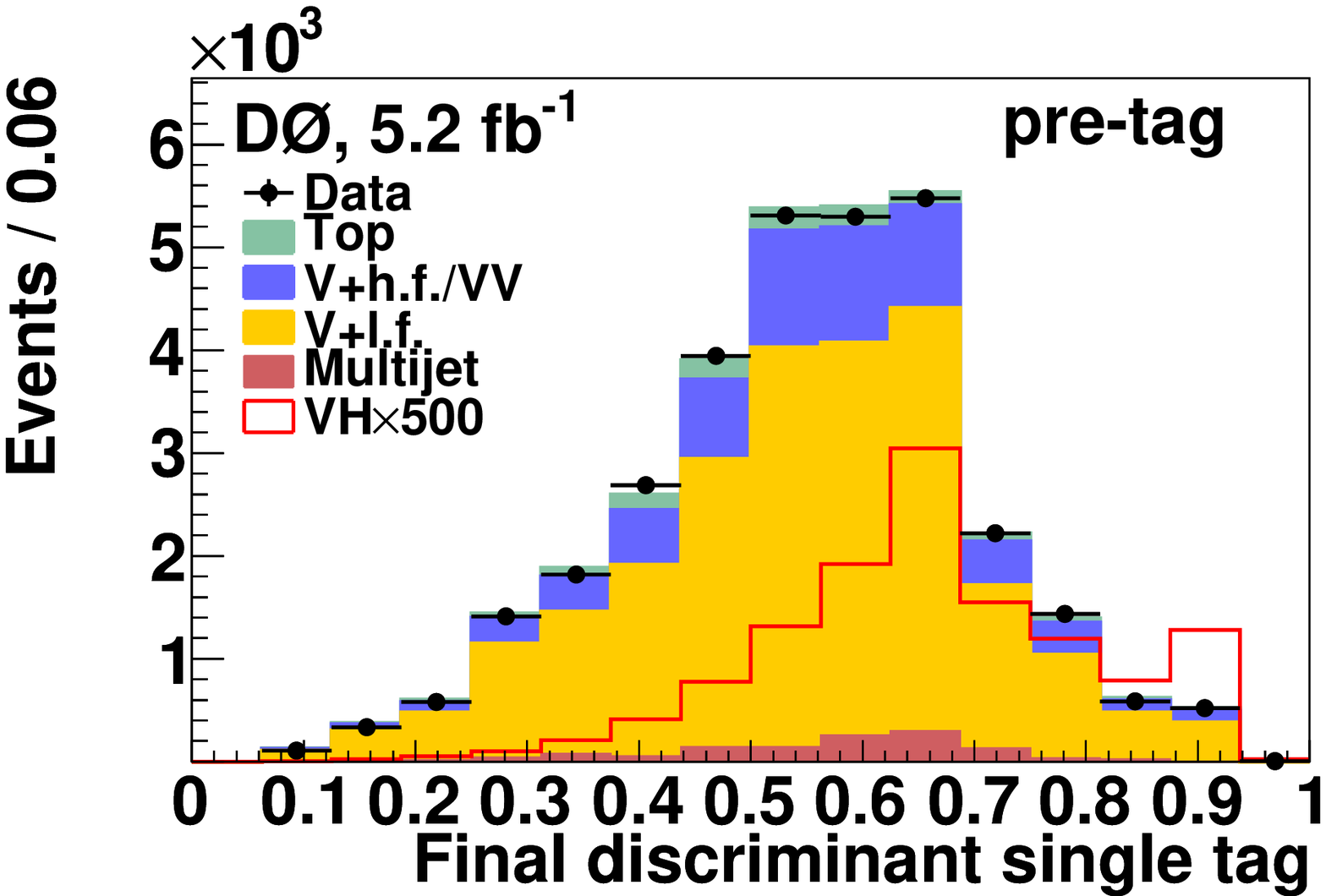}}
\subfigure[]{\includegraphics[width=8.5cm]{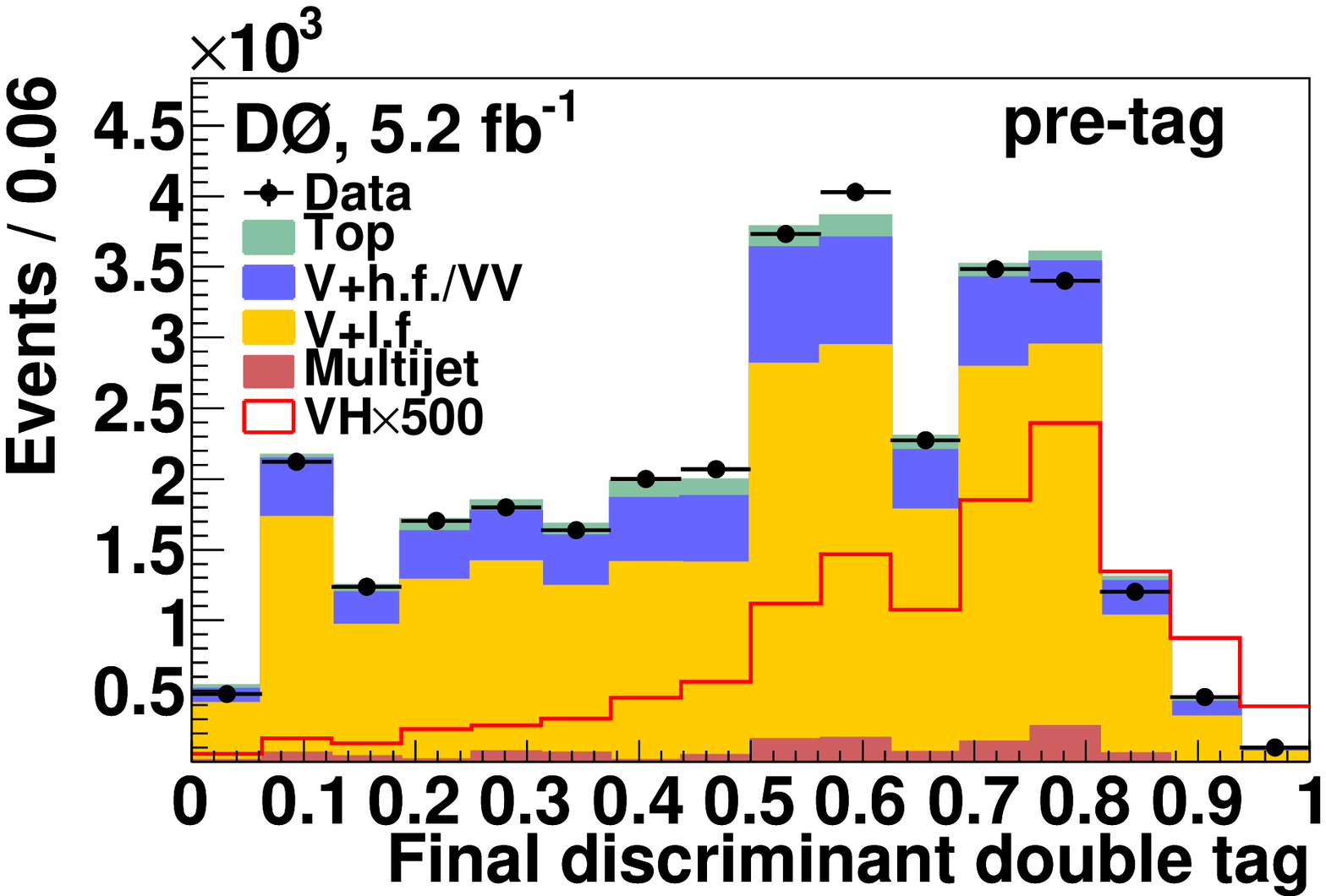}}
\caption{\label{PRL_DT_QCD}
Discriminants in the analysis sample:
(a) Multijet discriminant,
(b) Final single tag discriminant in the pre-tag sample,
(c) Final double tag discriminant in the pre-tag sample.
The signal includes $ZH$ and $WH$ production for $m_H=115$~GeV.
}
\end{figure*}

\begin{figure*}[htp]
\centering
\subfigure[]{\includegraphics[width=8.5cm]{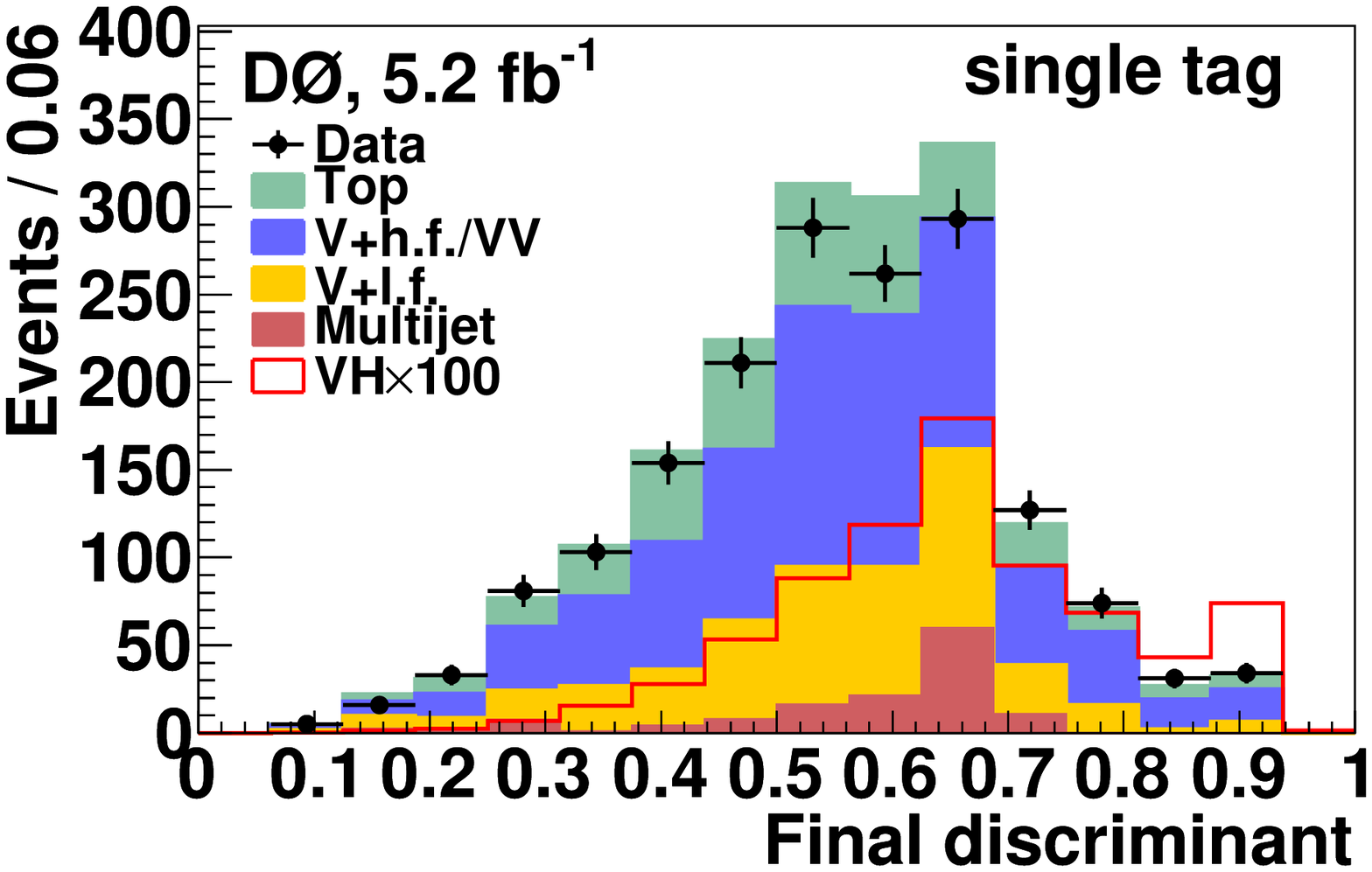}}
\subfigure[]{\includegraphics[width=8.5cm]{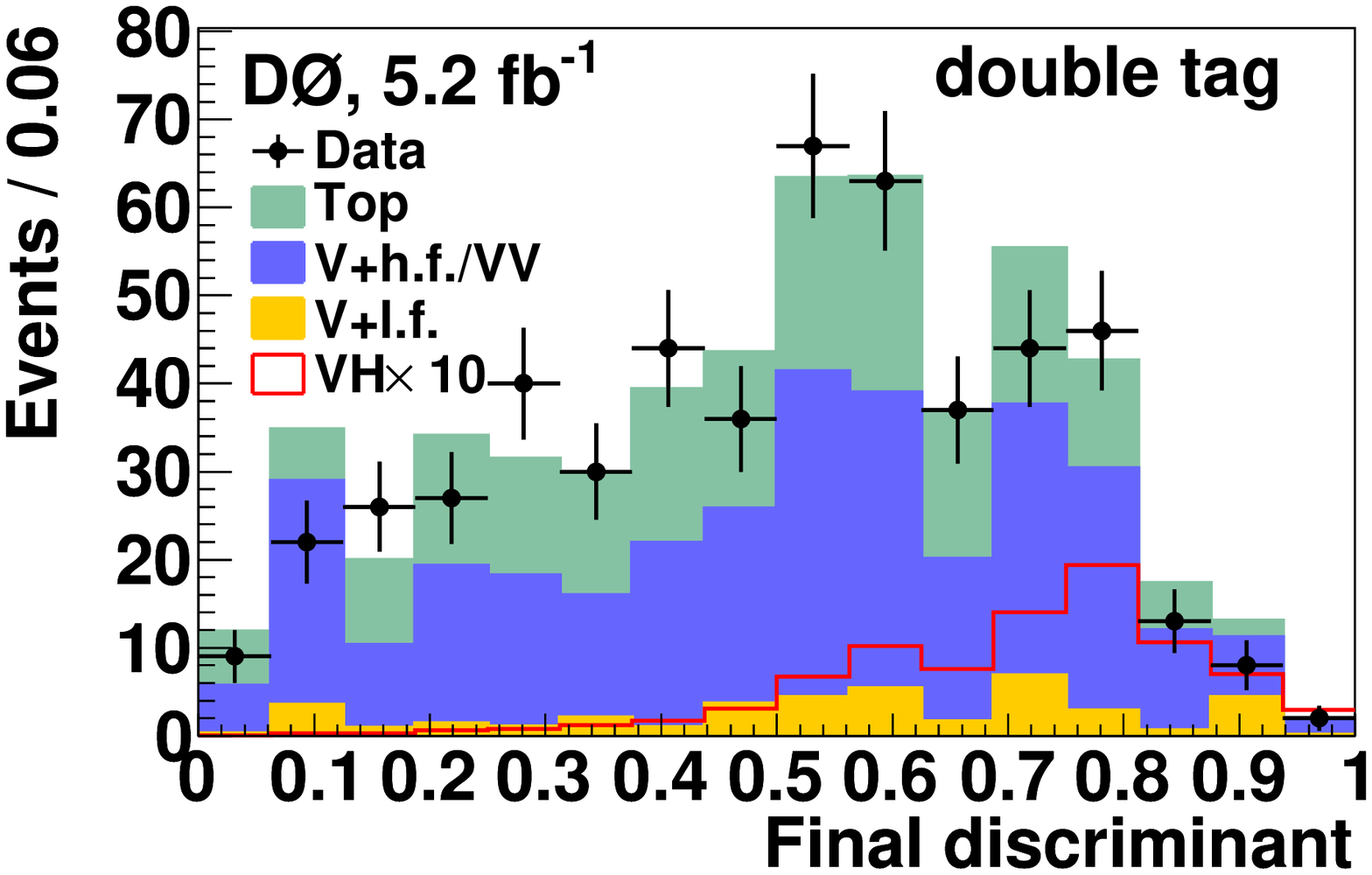}}
\caption{\label{prl_DT_Physics_1tag}
Discriminants in the analysis sample:
(a) Final single tag discriminant in the single tag sample,
(b) Final double tag discriminant in the double tag sample.
The signal includes $ZH$ and $WH$ production for $m_H=115$~GeV.
}
\end{figure*}

\begin{figure*}[htp]
\centering
\subfigure[]{\includegraphics[width=8.5cm]{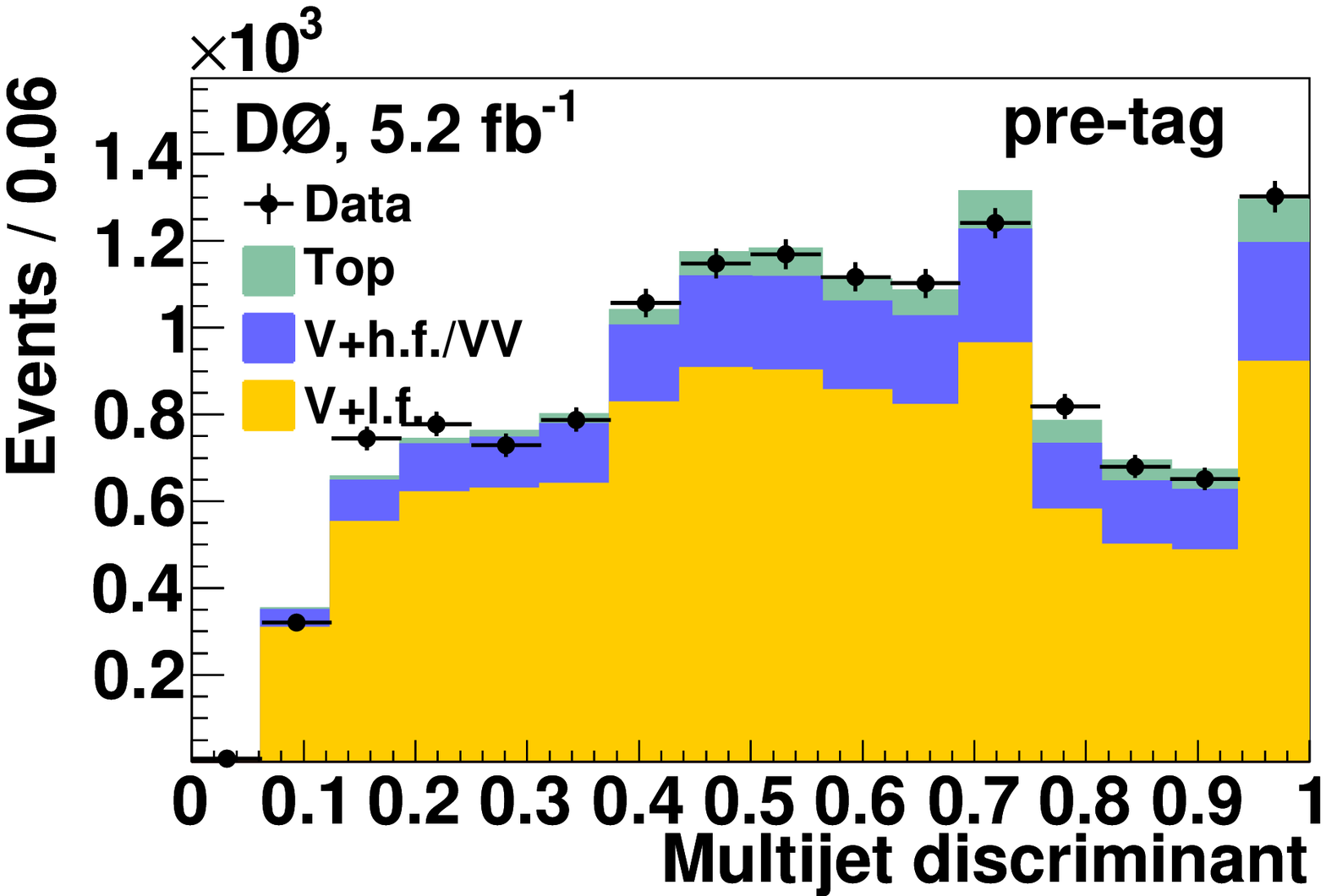}} \\
\subfigure[]{\includegraphics[width=8.5cm]{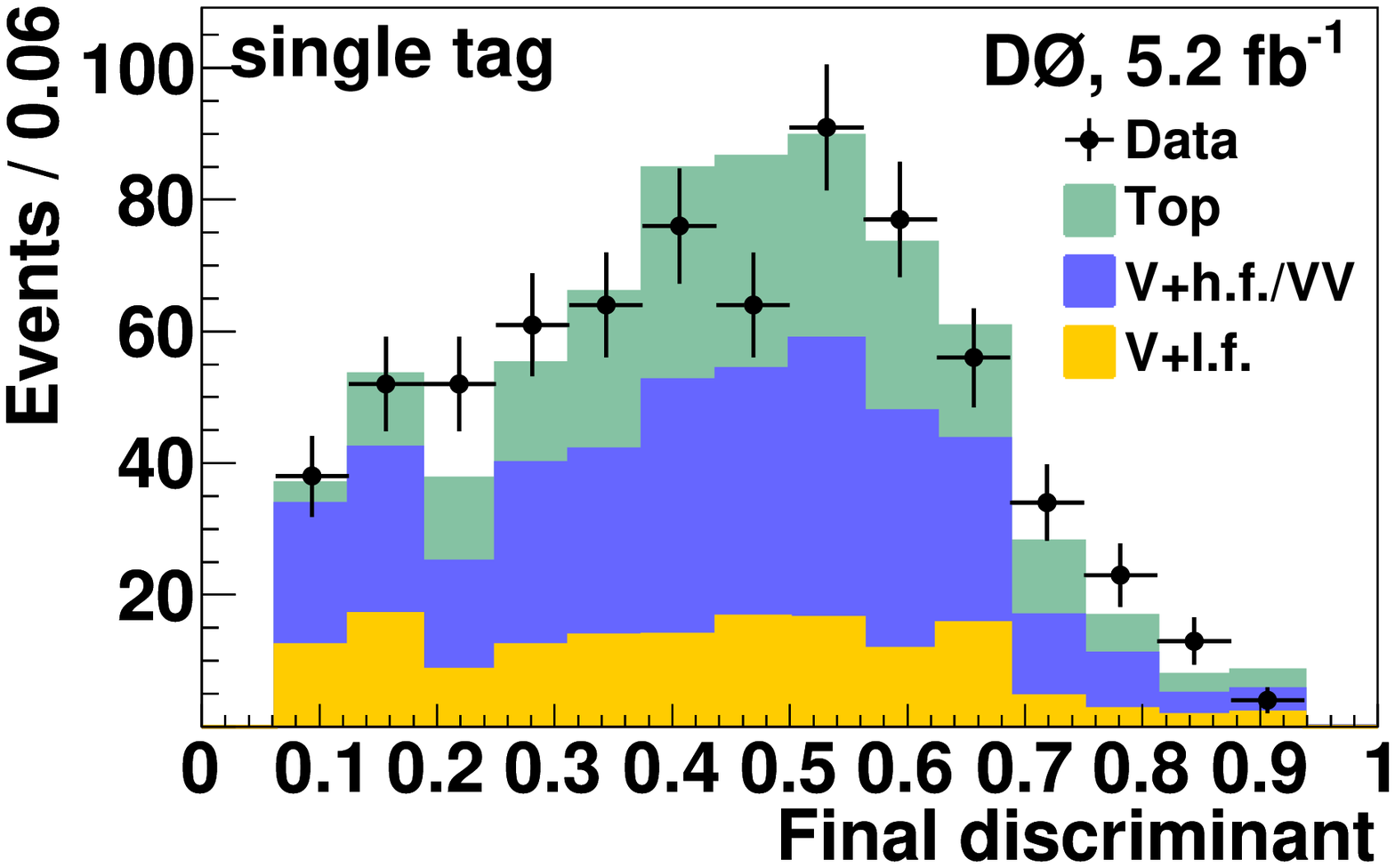}}
\subfigure[]{\includegraphics[width=8.5cm]{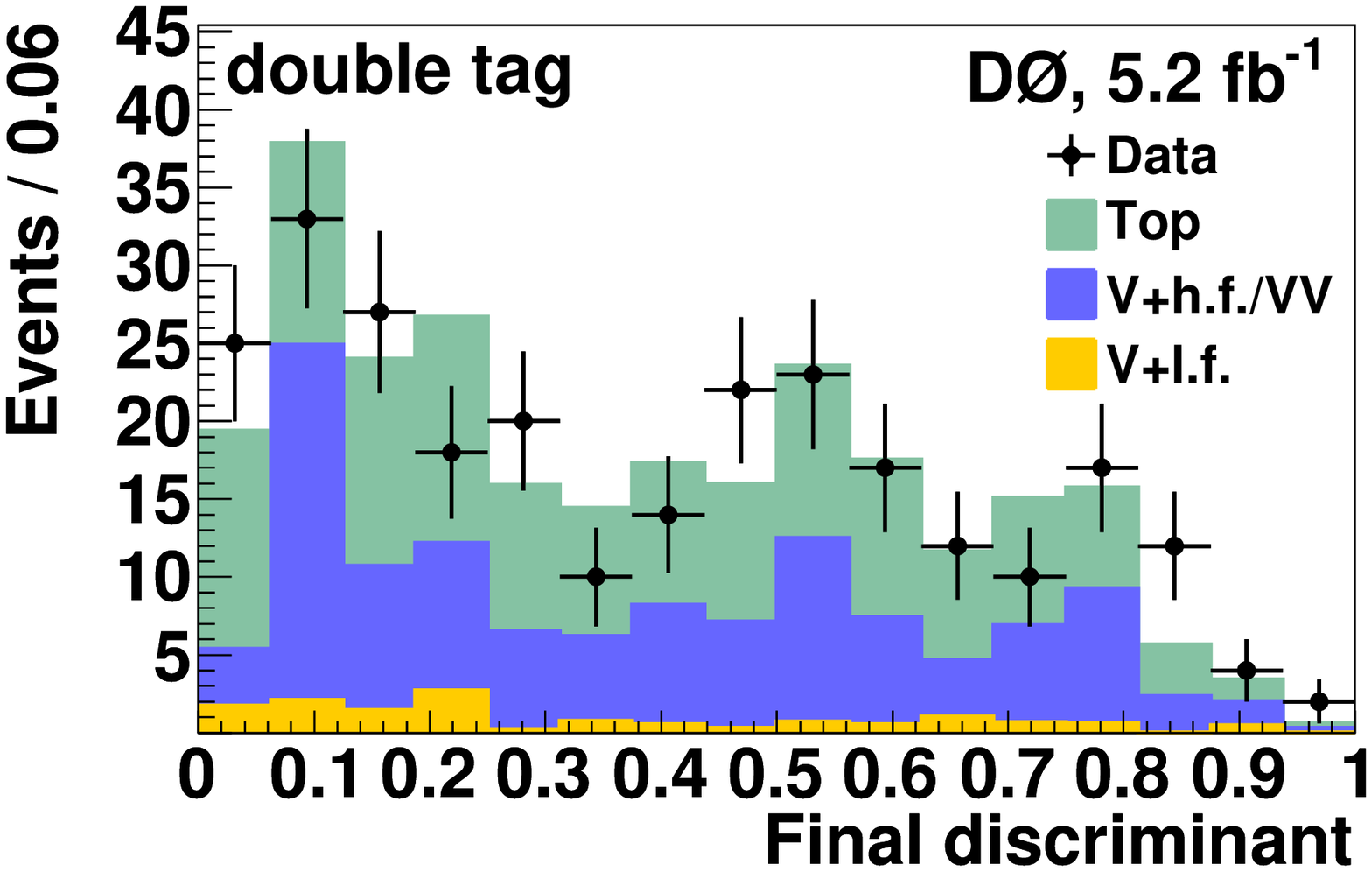}}
\caption{\label{prl_DT_Physics_1tag_ew1tag}  
Discriminants in the EW-control sample:
(a) Multijet discriminant,
(b) Final single tag discriminant in the single tag sample,
(c) Final double tag discriminant in the double tag sample.
}
\end{figure*}

\begin{figure*}[htp]
\centering
\subfigure[]{\includegraphics[width=8.5cm]{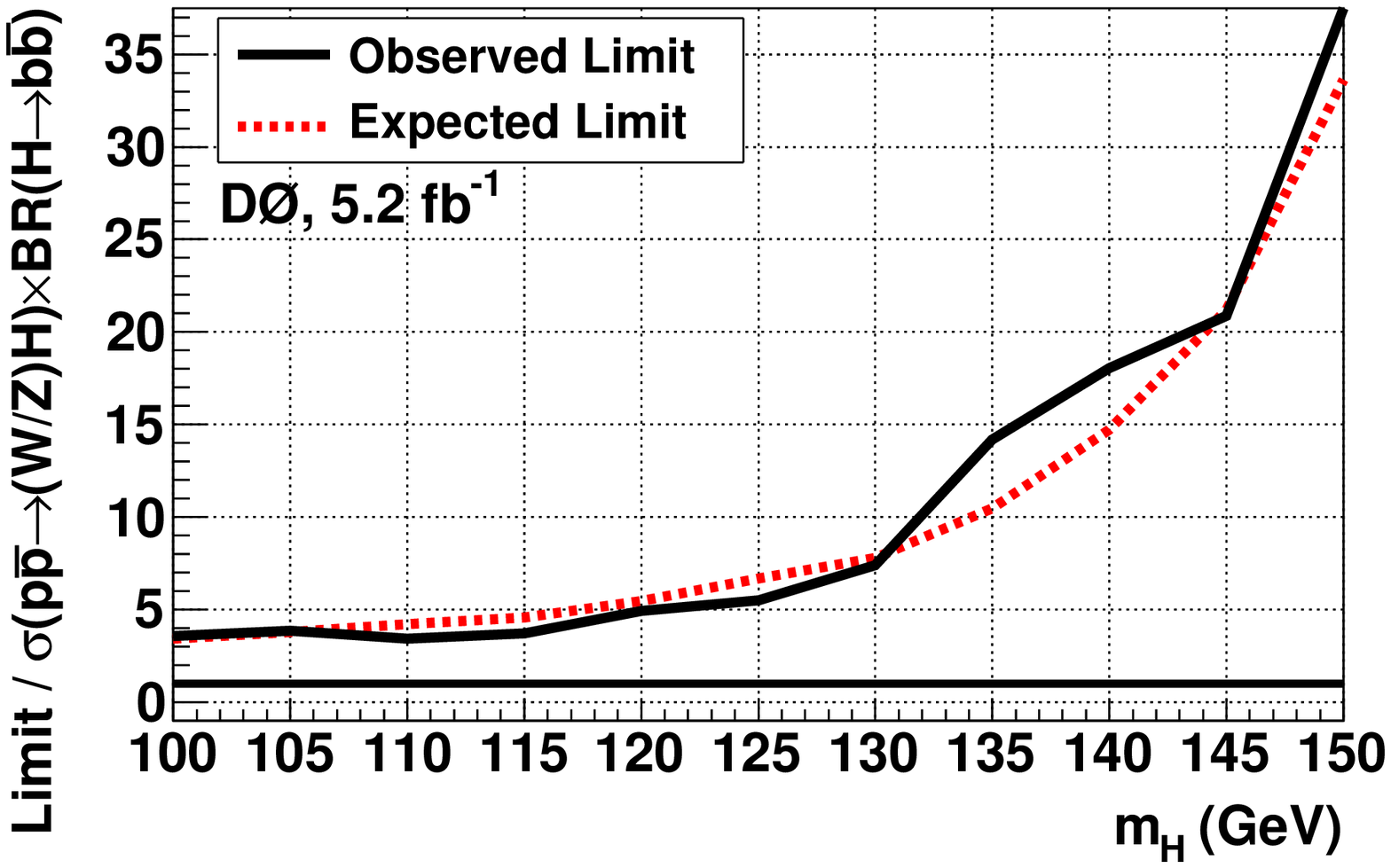}}
\subfigure[]{\includegraphics[width=8.5cm]{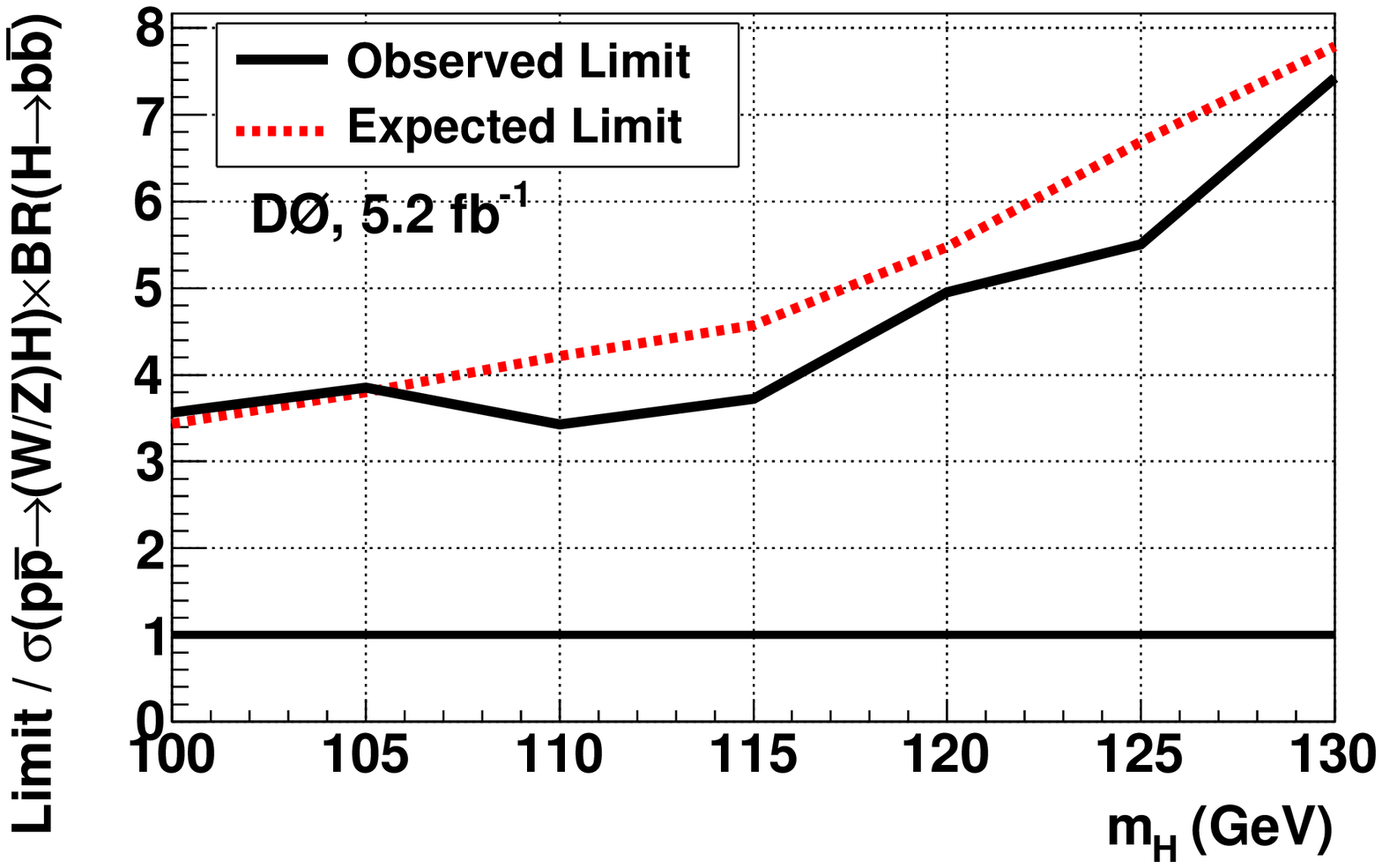}}
\caption{\label{LLR}
(a) Ratio of the observed (solid black) and expected (dotted red) 
exclusion limits to the SM production cross section multiplied by
branching fraction for $H\to\bbb$, as a function of $m_H$, 
(b) The same zoomed in the low mass region.
}
\end{figure*}


\begin{thebibliography}{99}
\bibitem{eta}
The pseudorapidity is defined as $\eta=-\ln[\tan(\theta/2)]$, 
where $\theta$ is the polar angle with respect to the proton beam direction.
\end{thebibliography}
\end{document}